\documentclass[twocolumn]{aastex63}

\usepackage{threeparttable}
\usepackage{comment}
\usepackage{hyperref}

\accepted{ApJ}
\shorttitle{Illuminating the dark side of cosmic star formation}
\shortauthors{Talia, Cimatti, Giulietti, Zamorani et al.}

\begin{document}

\title{Illuminating the dark side of cosmic star formation two billion years after the Big Bang}

\correspondingauthor{Margherita Talia}
\email{margherita.talia2@unibo.it}

\author[0000-0003-4352-2063]{Margherita Talia}
\affiliation{University of Bologna - Department of Physics and Astronomy  ``Augusto Righi" (DIFA), Via Gobetti 93/2, I-40129, Bologna, Italy}
\affiliation{INAF - Osservatorio di Astrofisica e Scienza dello Spazio, Via Gobetti 93/3, I-40129, Bologna, Italy}

\author[0000-0002-4409-5633]{Andrea Cimatti}
\affiliation{University of Bologna - Department of Physics and Astronomy, Via Gobetti 93/2, I-40129, Bologna, Italy}
\affiliation{INAF - Osservatorio Astrofisico di Arcetri, Largo E. Fermi 5, I-50125, Firenze, Italy}

\author[0000-0002-1847-4496]{Marika Giulietti}
\affiliation{SISSA, Via Bonomea 265, I-34136 Trieste, Italy}
\affiliation{University of Bologna - Department of Physics and Astronomy, Via Gobetti 93/2, I-40129, Bologna, Italy}

\author[0000-0002-2318-301X]{Gianni Zamorani}
\affiliation{INAF - Osservatorio di Astrofisica e Scienza dello Spazio, Via Gobetti 93/3, I-40129, Bologna, Italy}

\author[0000-0002-3915-2015]{Matthieu Bethermin}
\affiliation{Aix Marseille Univ, CNRS, LAM, Laboratoire d'Astrophysique de Marseille, Marseille, France}

\author[0000-0002-9382-9832]{Andreas Faisst}
\affiliation{IPAC, M/C 314-6, California Institute of Technology, 1200 East California Boulevard, Pasadena, CA 91125, USA}

\author[0000-0001-5891-2596]{Olivier Le F$\grave{\mathrm{e}}$vre}
\affiliation{Aix Marseille Univ, CNRS, LAM, Laboratoire d'Astrophysique de Marseille, Marseille, France}

\author[0000-0002-3893-8614]{Vernesa Smol$\check{\mathrm{c}}$i$\acute{\mathrm{c}}$}
\affiliation{Department of Physics, Faculty of Science, University of Zagreb\\
Bijeni$\acute{\mathrm{c}}$ka cesta 32, 10000 Zagreb, Croatia}




\begin{abstract} 
How and when did galaxies form and assemble their stars and stellar mass? 
The answer to these questions, so crucial to astrophysics and cosmology, requires the full reconstruction of the so-called cosmic star formation rate density (SFRD), i.e. the evolution of the average star formation rate per unit volume of the universe. 
While the SFRD has been reliably traced back to 10-11 billion years ago, its evolution is still poorly constrained at earlier cosmic epochs, and its estimate is mainly based on galaxies luminous in the ultraviolet and with low obscuration by dust. 
This limited knowledge is largely due to the lack of an unbiased census of all types of star-forming galaxies in the early universe.
We present a new approach to find dust-obscured star-forming galaxies based on their emission at radio wavelengths coupled with the lack of optical counterparts. Here, we present a sample of 197 galaxies selected with this method. These systems were missed by previous surveys at optical and near-infrared wavelengths, and 22 of them are at very high redshift (i.e. $z > 4.5$). The contribution of these elusive systems to the SFRD is substantial and can be as high as 40$\%$ of the previously known SFRD based on UV-luminous galaxies. 
The mere existence of such heavily obscured galaxies in the first two billion years after the Big Bang opens new avenues to investigate the early phases of galaxy formation and evolution, and to understand the links between these systems and the massive galaxies which ceased their star formation at later cosmic times.
\end{abstract}




\section{Introduction}\label{sec:intro}

How efficiently did gas transform into stars as a function of cosmic time? 
The answer to this key question requires to reconstruct the cosmic SFRD to the highest possible redshifts. 
However, despite the major progress achieved in the last decades in understanding galaxy evolution  \citep{madau2014}, several key questions remain open. 
The integration of the SFRD$(z)$ over redshift, making appropriate corrections for stellar evolution processes, yields the current stellar mass density $\rho_{\ast}(z)$. 
The results obtained so far  \citep{madau2014, Oesch2018} show a rather consistent picture up to $z\approx 3$. Several independent results indicate that the SFRD rapidly increases from $z = 0$ to $z \approx 1$, and flattens around $z \approx 2$ (the so-called ”cosmic noon”). 

However, major uncertainties remain at $z >3$  \citep{Casey2014, Magnelli2019}. 
At these high redshifts, it is still unclear whether the SFRD rapidly declines or remains rather flat  \citep{Gruppioni2013, RowanR2016, Novak2017, gruppioni2020}. 
The poor knowledge of the SFRD evolution at $z >3$ is primarily due to two main limitations. 
Firstly, the vast majority of SFRD estimates at these redshifts comes from the observation of Lyman-break galaxies (LBGs) in the rest-frame UV. 
This makes the results strongly dependent on the adopted dust extinction correction. 
Moreover, LBGs may not be fully representative of the whole population of the star-forming galaxies (SFGs) existing at these redshifts. 
Secondly, the available surveys in the IR (\emph{Spitzer}, \emph{Herschel}  \citep{lutz2011, magnelli2013}) and sub-mm/mm bands have either a limited sensitivity to galaxies at $z >3$ and/or are often plagued by large beam sizes which imply significant source blending. 
Indeed, since when the first deep surveys in the sub-mm regime (850--870 $\mu$m) uncovered the presence of sub-mm galaxies at mJy flux levels \citep[SMGs;][]{smail1997, hughes1998}, the coarse angular resolution of single-dish telescopes and the faintness, or sometimes complete lack, of the optical/near-infrared (OPT/NIR) counterparts posed serious challenges to their identification and characterization \citep[e.g.][]{dannerbauer2004, frayer2004}. 
In this respect, the intense starburst HDF 850.1 is a notable case because of the fifteen-years gap between its first discovery \citep{hughes1998} and the secure spectroscopic identification \citep{Walter2012}.
The ALMA follow up of increasingly larger samples of SMGs firstly identified with single-dish observations \citep[e.g.][]{simpson2014, simpson2017, dudzeviciute2020} proved fundamental to establish the physical properties of these bright ($S_{870\mu m}>$ 1--2 mJy) galaxies, which are typically located around z$\sim$2.5--3.0 \citep[e.g.][]{simpson2014, simpson2017, simpson2020, dudzeviciute2020}, are characterized by very high far-infrared (FIR) luminosities and SFRs \citep[several hundreds of M$_{\odot}$ yr$^{-1}$,][]{swinbank2014}, and constitute a significant fraction of the SFRD at cosmic noon. 
ALMA deep fields provide a complementary approach, typically reaching fainter flux limits of $S_{870\mu m}\sim$0.1--1 mJy, but they are currently too small to map sufficiently large volumes \citep[e.g.][]{walter2016, aravena2016, dunlop2017, franco2018, hatsukade2018}.
Different works report that a fraction of the ALMA-identified SMGs \citep[10--20$\%$; e.g.][]{franco2018, gruppioni2020, dudzeviciute2020} lack an OPT/NIR counterpart.
A favored explanation is that these UV-, or HST-dark galaxies, as they are often called, are extremely dust-obscured and/or lie at a much higher redshift than the bulk of the SMGs population ($>$4).
For example, \citet{wang2019} recently reported the results from the ALMA follow-up of a population of optically dark galaxies (H-dropouts), and confirmed a fraction of them to be massive dusty galaxies at high-redshift. They concluded that this population constitutes a significant fraction of the SFRD at high redshift ($>$3), but also left open the question of whether the fraction of currently missed SFGs might be even higher. 
A handful of extreme SFGs heavily obscured by dust and missed in OPT/NIR surveys has been indeed identified out to very high redshifts ($z\sim 5-6$) \citep{Dannerbauer2008, Wang2009, Walter2012, Riechers2013, Riechers2017, Marrone2018, riechers2020}.
However, it is unclear if these seemingly very rare objects are representative of a more vast and elusive population, and whether their high luminosities are partly amplified by gravitational lensing due to intervening matter along the line of sight \citep{ciesla2020, bakx2020}. 
To answer this question, several efforts are being pursued to uncover dusty systems at very high redshifts ($>$3--4) with a combination of space- and ground-based data in the FIR to mm spectral range \citep{Ivison2016, Casey2018, Duivenvoorden2018, Magnelli2019, lefevre2019, bethermin2020, faisst2019}.

In this work, we present the results of a search for dusty UV-dark galaxies at $z \gtrsim 3$ selected at radio frequencies, taking advantage of the depth and area of the VLA-COSMOS 3 GHz Large Project \citep{Smolcic2017}. 
SFGs display radio emission due to a combination of synchrotron radiation emitted by electrons accelerated in supernova remnants and free-free continuum from H II regions. 
As a consequence, the radio luminosity of SFGs is an indicator of the SFR, provided that the contamination from AGN emission is negligible. 
As radio photons are immune to dust extinction, it is therefore possible to exploit deep radio surveys to search for dust-obscured, highly star-forming systems at high redshifts. 
Moreover, the angular resolution of interferometric data allows us to localize the OPT/NIR counterparts much more reliably than for sources selected in the FIR/mm where the beam size and the consequent blending of different sources can be a severe limitation. 
The selection in the radio can be advantageous with respect to FIR/sub-mm surveys also because it is independent of dust temperature. 
It has been suggested that the dust temperature (T$_{dust}$) correlates with the infrared luminosity, specific star formation rate and redshift \citep{Schreiber2018, bethermin2015, faisst2017}. 
For instance, the average T$_{dust}$ almost doubles from $z \approx 0$ to $z \approx 4$.
This would imply that FIR-to-mm surveys could be affected by selection biases which depend on T$_{dust}$, and hence the wavelength at which the grey-body emission peaks; instead, radio emission is free from these effects.
The nature of the T$_{dust}$-redshift relation is indeed still debated \citep[e.g.][]{dudzeviciute2020, faisst2020}, with recent works suggesting that the observed trend could be mostly due to the relatively high median SFR of the current sample of dusty-SFGs at $z>5$ \citep{riechers2020}.

We give magnitudes in the AB photometric system, assume a  \citet{chabrier2003} initial mass function, and use the following cosmological parameters: $\Omega_{M}$=0.3, $\Omega_{\Lambda}$=0.7, $H_{0}$=70 km s$^{-1}$ Mpc$^{-1}$.

\section{Radio selection of dust-obscured systems}  \label{sec:sample_sel}

Our study relies on the data-set collected with the VLA-COSMOS 3 GHz Large Project \citep{Smolcic2017}, a survey based on 384-hour observations of the COSMOS field (2 deg$^2$) with the Karl G. Jansky Very Large Array (VLA) interferometer at 3 GHz ($\lambda=10$ cm). 
This survey already proved to be deep enough to allow the identification of SFGs at $z > 3$ \citep{Novak2017}, but our aim is to go beyond the results obtained so far and explore the existence of dusty galaxies that were missed in previous studies. 

\begin{figure*}[!ht]
	\centering
	\includegraphics[scale=0.8, clip=true]{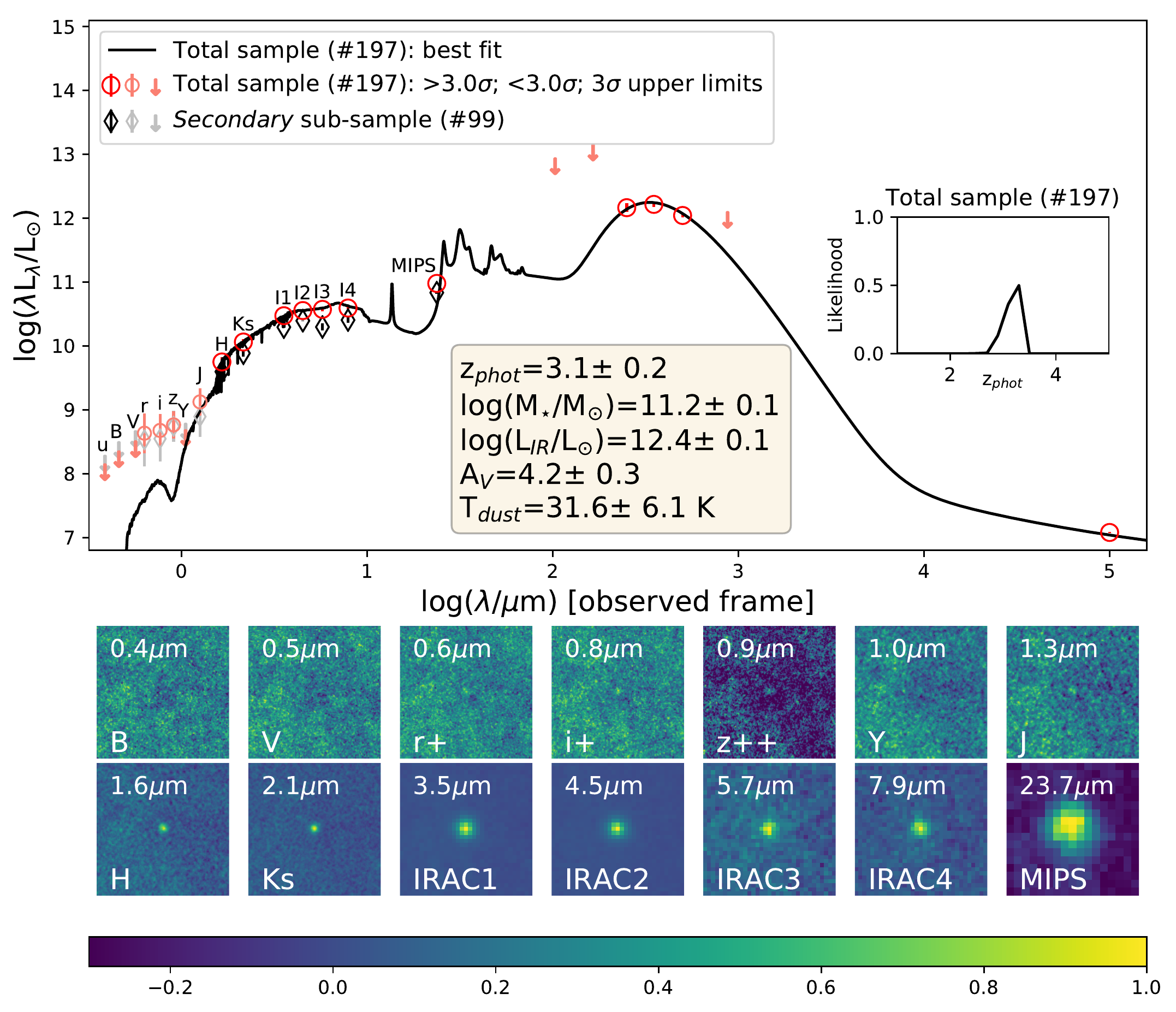} 
	\caption{\emph{Top:} median SED of the total sample of the 197 radio sources analyzed in this work (red and salmon points).
Error bars are plotted for each data-point, although in some cases they are smaller than the points.
The black line is the best-fit from \texttt{MAGPHYS}.
Detections and 3$\sigma$ upper limits are derived from stacked images up to 24 $\mu$m, while median fluxes from the \emph{Super-deblended} catalog \citep{Jin2018} are used in the FIR regime. 
The inset shows the likelihood distribution of the photometric redshift for the total sample of 197 galaxies.
We also show the median OPT-to-MIR SED of the \emph{secondary} sub-sample of 99 galaxies (black and grey symbols). 
\emph{Bottom}: normalized to the peak flux OPT-to-MIR stacked images of the total sample. }
	\label{fig:medsed}
\end{figure*}

We built our sample of radio-selected, UV-dark galaxies following these steps:

\begin{enumerate}
\item Starting from the VLA source catalog \citep{Smolcic2017}, we selected a parent sample of 8850 objects with flux density at 3 GHz (10 cm) $S_{\rm 3 GHz}> 12.65$ $\mu$Jy beam$^{-1}$, corresponding to 5.5$\sigma$, which is the threshold over which the estimated fraction of spurious sources over the entire catalog is only 0.4$\%$ \citep{Smolcic2017}.
\item We subsequently excluded all sources that were located inside masked or bad areas of the UltraVISTA footprint \citep{laigle2015}. The total effective area used in our analysis is 1.38 deg$^2$, inside which we counted 5982 sources above the S/N cut. 
\item We excluded all multi-component radio sources as identified by  \citet{Smolcic2017}, because the radio emission in these sources is likely associated with AGN activity, rather than to star-formation.
\item Then, we cross-correlated our catalog of radio sources with the COSMOS2015 photometric catalog \citep{laigle2015} within a search radius of 0.8$\arcsec$ \citep{Smolcic2017}. 
We identified 476 VLA sources without a COSMOS2015 counterpart.
\item Finally, we restricted our selection to isolated sources, in order to avoid the risk of contamination of the photometry of the galaxies in our sample by nearby, physically unrelated objects, since source blending would make photometry and identification of multi-wavelength counterparts uncertain.
In particular, we visually inspected the so-called $\chi^2$-image used as the detection map to build the COSMOS2015 catalog \citep{laigle2015}, which is a map obtained by co-adding the $z++$, $Y$, $J$, $H$ and $Ks$ images, and we excluded from our final sample all sources whose radio 3$\sigma$ isophote intersects the NIR 3$\sigma$ isophote of nearby objects in the $\chi^2$-image. 
We did not find strong evidence of gravitational lensing effects.
\end{enumerate}

\noindent The final sample analyzed in this work counts 197 galaxies.

\section{Stacking analysis and possible AGN contribution} \label{sec:stacked_SED}

Our first approach was to statistically analyze the properties of these galaxies, because of their extreme faintness at OPT/NIR wavelengths (down to limiting magnitudes AB = 24.0-24.7 in the NIR bands). 
We built the median SED of our total sample of 197 galaxies (Fig. \ref{fig:medsed}) by stacking the COSMOS images in each photometric filter, from the optical to 24 $\mu$m, at the radio positions. 
In particular, in the optical regime we used the same maps as in  \citet{laigle2015} (see their Table 1 for a summary), in the NIR bands the UltraVISTA DR3 maps, and in the MIR (mid-infrared) regime the SPLASH/IRAC, and MIPS maps (see next section).
In the FIR regime we did not stack directly Herschel and SCUBA maps. Instead, we computed the median of the flux in each photometric band from the \emph{Super-deblended} catalog \citep{Jin2018} and we employed survival analysis \citep{feigelson1985} to properly account for the upper-limits for the undetected sources. 

Quite remarkably, even in the stacked images, which have an effective depth $\sim$14 times higher than the original maps, there are only marginal (slightly more than 2$\sigma$) detections in the \emph{r+}, \emph{i+},  \emph{z++}, and \emph{J} bands, at the current depths, while significant fluxes emerge only at longer wavelengths, starting from the H-band.

We fitted the broad-band photometry in the entire wavelength range up to radio with the \texttt{MAGPHYS} code \citep{dacunha2008, battisti2019}.
\texttt{MAGPHYS} is a physically-motivated model package which self-consistently models the OPT-to-radio emission of a galaxy.
In particular, the emission from stellar populations in galaxies is computed using the  \citet{bruzual2003} models, with delayed exponentially declining star-formation histories, together with random bursts superimposed onto the continuous model, with a range of ages and exponential timescale values. 
The effects of dust attenuation are included as prescribed by  \citet{charlot2000}. 
The models are uniformly distributed in metallicity between 0.02 and 2 times solar.
The emission from dust accounts both for the dust emission originating from the stellar birth clouds and for the dust emission originating from the ambient (i.e. diffuse) ISM \citep{dacunha2008}. 
Radio emission is also included as prescribed by \citet{daCunha2015}.
The optical and infrared libraries are linked together via energy balance.
 The broad wavelength range used in the analysis helps in mitigating the degeneracy between stellar age, dust and photometric redshift \citep{daCunha2015}.

We derived a photometric redshift for the median SED of the total sample of 197 galaxies of $z_{\rm med}=$ 3.1 $\pm$ 0.2, which is slightly higher than the typical median redshift of the general population of ALMA-identified SMGs \citep[z$\sim$2.6$\pm$0.1; e.g.][]{simpson2017}, but comparable to the median redshift of K-dark SMGs in the AS2UDS survey \citep[$z=3.0\pm$0.1][]{dudzeviciute2020}\footnote{\citet{dudzeviciute2020} report that 17$\%$ of their sample of bright SMGs is K-dark. We point out that the limiting K-band magnitude in the AS2UDS field is slightly deeper than in the UltraVISTA DR4, namely $AB$ = 25.1 (5$\sigma$).}.
We summarize the other median properties of the sample, as derived from the stacked SED, in Table \ref{tab1}.

The uncertainties quoted in Fig. \ref{fig:medsed} and Table \ref{tab1} take into account the fact that, at our radio flux threshold, $\sim$0.4$\%$ of spurious sources are expected \citep{Smolcic2017}. 
Under the assumption that all spurious sources inside the sampled area were picked up by our selection, we estimated that $\sim$10 out of the 197 sources which contributed to the stacked SED, might be spurious. 
We performed 100 realizations of the stacking analysis by substituting 10 random sources and we added quadratically the standard deviation of the distributions of the output properties to the \texttt{MAGPHYS} errors from the main fit.

The median $L_{IR}$ (2.3 $\times$ 10$^{12}$ L$_{\odot}$) is in the so-called ultra-luminous infrared galaxy (ULIRG) regime.
The effective T$_{dust}$\footnote{T$_{dust}$ derived by \texttt{MAGPHYS} is an average luminosity-weighted value obtained by the fit of the multiple dust components.} is 31.6$\pm$6.1 K. 
We also derived it by using a different approach, i.e. by fitting a modified black-body with $\beta$ = 1.5 to the median stacked FIR photometry. We obtain T$_{dust}$ = 33$\pm$4 K, which is perfectly in line with the \texttt{MAGPHYS} estimate.
Our median T$_{dust}$ is slightly colder, but broadly consistent, with that expected from the redshift evolution in Main Sequence galaxies from the literature (T$_{dust}$ = 42$\pm$3 K at $z=$3.1)\footnote{We quote the light-weighted average T$^{light}_{dust}$, obtained by applying eq. 6 from \citet{Schreiber2018}. This is comparable to the temperature one would measure by using a modified blackbody model with a single temperature and an emissivity of $\beta$ = 1.5.} \citep{Schreiber2018}.

The inferred high dust extinction ($A_V\sim$ 4.2 mag) confirms the strong obscuration of these galaxies.
It is also interesting to notice that the median values of SFR$_{IR}$ and $\mathcal{M}_{\star}$ lie within 0.3 dex of the star formation Main Sequence at z$\sim$3 \citep{rodighiero2011, speagle2014, tasca2015, talia2015}.

The IR-based and radio-based SFR estimates are in very good agreement. 
The $q_{TIR}$ parameter (IR-radio correlation) is consistent with the results ($q_{TIR}$=2.28 at $z$=3.1) obtained for a sample of radio-selected SFGs \citep{Novak2017}, suggesting the lack of strong AGN activity in the radio band \citep{delhaize2017}. 

\begin{table}
	\caption{Summary of the median physical properties of the total sample of 197 galaxies analyzed in this work}
	\centering
    	\label{tab1}
	\begin{center}
    	\begin{tabular}{l l l}
    	\hline \hline

	\multicolumn{3}{l}{} \\
	\multicolumn{3}{l}{From SED fitting to the median stacked photometry$^{a}$}\\
	\hline
    	z$_{phot}$		& 			& 3.1$\pm$ 0.2         			\\
    	M$_{\star}$ 		& M$_{\odot}$		& (1.7$\pm$0.3)$\times$ 10$^{11}$	\\
    	L$_{IR}$$^{b}$		& L$_{\odot}$		& (2.3$\pm$0.5)$\times$ 10$^{12}$	\\
   	A$_V$			& mag			& 4.2$\pm$0.3				\\
   	T$_{\rm dust}$		& K			& 31.6$\pm$6.1				\\

	\multicolumn{3}{l}{} \\
	\multicolumn{3}{l}{Median value from catalog}\\
	\hline
	S$_{3 GHz}$		& $\mu$Jy		& 18.2$\pm$0.5			\\

	\multicolumn{3}{l}{} \\
	\multicolumn{3}{l}{Derived quantities$^{a}$}\\
	\hline
	L$_{2-10 keV}$$^{c}$ 	& erg s$^{-1}$ 		& (2.3$\pm$0.7)$\times$ 10$^{42}$	\\
	L$_{1.4 GHz}$$^{d}$	& erg s$^{-1}$ Hz$^{-1}$& (1.7$\pm$0.1)$\times$ 10$^{31}$	\\
	$q_{TIR}$		& 			& 2.13$\pm$0.13	 			\\
   	SFR$_{IR}$$^{e}$	& M$_{\odot}$ yr$^{-1}$	& 309$\pm$70				 \\
	SFR$_{rad}$$^{f}$	& M$_{\odot}$ yr$^{-1}$	& 328$\pm$8	 		 	\\
	SFR$_{Xray}$$^{e}$	& M$_{\odot}$ yr$^{-1}$	& 398$\pm$115			 	\\

    	\hline \hline
    	\end{tabular}
	\begin{tablenotes}
	$^{a}$ Uncertainties on SED fitting-derived properties are the average of 16th and 84th percentiles of the parameters distributions. Uncertainties on derived properties come from the propagation of errors on fluxes; \\
	$^{b}$ L$_{3-1100\mu m}$; \\
	$^{c}$ derived from X-ray stacking and assuming \citep{civano2016} a power-law photon index $\Gamma$=1.8; \\
	$^{d}$ derived from $S_{3 GHz}$ assuming \citep{Novak2017} a spectral index $\alpha$=-0.7; \\
	$^{e}$ derived following  \citet{kennicutt2012}, scaled to a  \citet{chabrier2003} IMF; \\
	$^{f}$ derived following  \citet{Novak2017} (see their eq. 13).
	\end{tablenotes}
	\end{center}
\end{table}

In order to investigate further the possible presence of hidden AGN activity in our sample, we also performed X-ray stacking. 
We used the publicly-available CSTACK\footnote{CSTACK was created by Takamitsu Miyaji and is available at \url{http://lambic.astrosen.unam.mx/cstack/}} tool to stack Chandra soft ([0.5--2] keV) X-ray images from the Chandra-LEGACY survey \citep{civano2016} at the radio position of the objects in our sample.
We excluded from the stack one source that has a counterpart within 1$^{\prime\prime}$ in the Chandra-LEGACY point-source catalog. 
The stacked count rate detection, at 3.4$\sigma$ significance, was converted into a stacked 2--10 keV luminosity by assuming a power-law spectrum with a slope \citep{luo2017} $\Gamma$ of 1.8.
The X-ray based SFR is somewhat higher than, but consistent with, the IR- and radio-based estimates (Table \ref{tab1}), suggesting that, on average, star-formation alone is enough to produce the observed $L_{2-10keV}$. 
We point out that our estimate of the SFR$_{Xray}$ is only a lower limit, because we did not consider the effects of an intrinsic absorption. For a column density of N$_{H}$$\sim$10$^{22}$cm$^{-2}$, typical of massive (10$^{10}-$10$^{11}$M$_{\odot}$) SFGs \citep{buchner2017}, the SFR$_{Xray}$ would be a factor 1.3 higher, not altering significantly our conclusions.

Summarizing, the stacking analysis gave us important information at statistical level on the nature of our sample of UV-dark galaxies: in particular it highlights that the bulk of the population is constituted by extremely dust-obscured galaxies at $z\sim$ 3, which were unaccounted for, up to now, by surveys based on selections at OPT/NIR observed wavelengths. 

\section{Analysis of individual sources} 

\subsection{The multi-wavelength catalog.} \label{sec:catalog}

We tried to go beyond this statistical analysis and to investigate more in detail the nature of these galaxies on an individual basis.
To achieve this scope, we exploited a multiwavelength catalog assembled from the deepest maps and catalogs available in the COSMOS field.
We searched for counterparts to each radio source of our sample in the following catalogs and/or data-sets.

\begin{itemize}
\item The latest public data release catalog (DR4) of the UltraVISTA survey \citep{mccracken2012}, where the $Ks$-band data reach a limiting 5$\sigma$ magnitude $AB=24.5-24.9$, fainter than the $AB=24.0-24.7$ limits of the DR2 images used for the COSMOS2015 catalog \citep{laigle2015}. 
\item IRAC (channels 1-4) fluxes from the v2.0 mosaics of the \emph{ Spitzer} Large Area Survey with Hyper-Suprime-Cam \citep{steinhardt2014} (SPLASH\footnote{\url{http://splash.caltech.edu/}}), extracted inside an aperture with radius of 2.9$\arcsec$. In particular, fluxes were extracted blindly with the code \texttt{SEXTRACTOR} \citep{bertin1996} and then matched with the radio positions using a search radius \citep{Smolcic2017} of 1.7$\arcsec$.  
We estimated the fraction of possible false associations of our radio sources with IRAC counterparts in the following way. We constructed a sample of $\sim$170,000 mock sources in empty regions of the radio map, but close enough (within 60$\arcsec$) to our 197 sources in order to sample regions with similar IRAC depth. We excluded mock sources with a counterpart in the COSMOS15 catalogue within 1.7$\arcsec$, in order to mimic the 5th step of our selection (Sec. \ref{sec:sample_sel}) aimed at excluding radio galaxies potentially contaminated by nearby bright sources\footnote{The average separation between the not-isolated UV-dark radio-galaxies and the COSMOS15 possible contaminant is 1.7$\arcsec$.}.
Finally, we matched the mock sources to the IRAC catalogues in the same way as the real radio data and we found a 0.7$\%$ percentage of spurious associations.
\item MIPS 24 $\mu$m fluxes from the latest release (G03) of the SCOSMOS data \citep{sanders2007}.
\item The \emph{Super-deblended} catalog \citep{Jin2018}, which includes photometry from Spitzer, Herschel, SCUBA, AzTEC, and MAMBO using as priors the radio positions at 3 GHz.
\item \emph{A3COSMOS} catalog \citep{liu2019}, which includes photometry at sub-mm wavelengths from the ALMA archive. In particular, we searched for counterparts to our sources in the version \texttt{v.20180801} of the catalog within a radius of 0.8$\arcsec$.
\item 1.4 GHz catalog from the VLA-COSMOS survey \citep{schinnerer2010}.
\end{itemize}

We find that only $\sim$10$\%$ of the UV-dark radio galaxies in our total sample have an observed $S_{850\mu m} \geq$ 4 mJy ($>$ 3$\sigma$), which is the usual flux threshold to define bright SMGs in single-dish surveys. 
We point out that the flux densities at 850 $\mu m$ for our sample come mainly from super-deblended photometry of SCUBA-2 sources (we find only 14 ALMA counterparts in the A3COSMOS catalogue). 
An ALMA follow-up of all our UV-dark galaxies would be fundamental to assess the actual overlap of our radio selection with the general population of SMGs.

Thanks to our extended dataset we could build the photometric NIR(MIR)-to-radio SED of 98 individual sources for which we could collect at least one FIR and one NIR-to-MIR photometric point (our \emph{primary} sub-sample). 
We point out that all galaxies with an IRAC counterpart in this sample, (all but one, with an estimated z$_{phot}\sim$0.5), have radio-IRAC separations below 1.3$\arcsec$. 
From our simulations we estimated that within such a radius the percentage of potentially spurious associations between the radio and IRAC bands is 0.1$\%$.

On average we had significant detections in at least $\sim$6-7 filters from NIR to FIR per galaxy, not counting the radio band at 3 GHz, while for only $\sim$3$\%$ of the \emph{primary} sub-sample we could collect only two photometric points (plus upper limits), which allowed us, however, to derive an estimate of their redshift and $L_{IR}$. 

We stress that the definition of \emph{dark} galaxies is not absolute. Specifically, in our case it is related to the lack of counterparts for our radio-selected galaxies in the COSMOS15 catalog. By taking advantage of deeper data which became available since the COSMOS15 release, together with the availability of precise positions from the radio data, we were nonetheless able to assign a faint NIR counterpart to a fraction of our galaxies. 
However, since the counterpart identification of the VLA-COSMOS 3 GHz general sample was based on the COSMOS15 catalog \citep{Smolcic2017, Novak2017}, not on the deeper UltraVISTA maps, our sources were not considered in previous works for the determination of the cosmic SFRD.

We found that 24 radio sources do not have a counterpart in any NIR-to-FIR bands (we only consider detections at a 3$\sigma$ significance level). 
For the remaining 75 sources, though being detected in at least one NIR-to-FIR band, we could not collect enough photometric points to model the observed SED. 
These latter two groups of galaxies constitute our \emph{secondary} sub-sample.

\begin{table}
	\caption{Median$^{a}$ of the distribution of physical properties of the individual galaxies belonging to the \emph{primary} sub-sample}
	\centering
    	\label{ext_tab1}
	\begin{center}
    	\begin{tabular}{l l l}
    	\hline \hline

	\multicolumn{3}{l}{} \\
	\hline
    	z$_{phot}$		& 			& 3.1$\pm$0.1\\
    	M$_{\star}$ 		& M$_{\odot}$		&  (2.0$\pm$0.2)$\times$ 10$^{11}$\\
    	L$_{IR}$$^{b}$		& L$_{\odot}$		&  (3.2$\pm$0.4)$\times$ 10$^{12}$\\
   	A$_V$			& mag			&  4.2$\pm$0.1\\
   	T$_{\rm dust}$		& K			&  39.5$\pm$0.7\\
	S$_{3 GHz}$		& $\mu$Jy		& 18.4$\pm$0.7\\
	L$_{1.4 GHz}$$^{c}$	& erg s$^{-1}$ Hz$^{-1}$& (1.7$\pm$0.2)$\times$ 10$^{31}$	\\

    	\hline \hline
    	\end{tabular}
	\begin{tablenotes}
	$^{a}$ For each parameter we quote the median value of the distribution. The associated errors have been estimated using the median absolute deviation (MAD), defined \citep{hoaglin1983} as MAD $=1.482\times median(|x_{i}-median(x_{i})|)$, divided by $\sqrt N$, where \emph{N} is the number of \emph{i} objects.; \\
	$^{b}$ L$_{3-1100\mu m}$; \\
	$^{c}$ derived from $S_{3 GHz}$ assuming \citep{Novak2017} a spectral index $\alpha$=-0.7.
	\end{tablenotes}
	\end{center}
\end{table}

\subsection{Physical properties and very high-redshift candidates} \label{sec:individual_SEDs}

For 98 out of 197 galaxies we could collect at least one FIR and one NIR-to-MIR photometric points and partially reconstruct the NIR-to-radio individual SED.
We call this group of 98 sources the \emph{primary} sub-sample, and the remaining 99 sources the \emph{secondary} sub-sample.
We used again the \texttt{MAGPHYS} code to perform SED fitting and derive the photometric redshifts and physical properties of the individual galaxies belonging to the \emph{primary} sub-sample.
In the optical regime we assumed the upper limits from  \citet{laigle2015}, while in the NIR-to-MIR bands we derived the upper limits for each undetected source directly from the maps.
Uncertainties on the photometric redshifts were derived from the 16th and 84th percentiles of the probability distribution produced by \texttt{MAGPHYS}. 
This accounts for uncertainty in the photometry as well as on the model galaxy templates.

\begin{figure}[]
	\centering
	\includegraphics[scale=0.6]{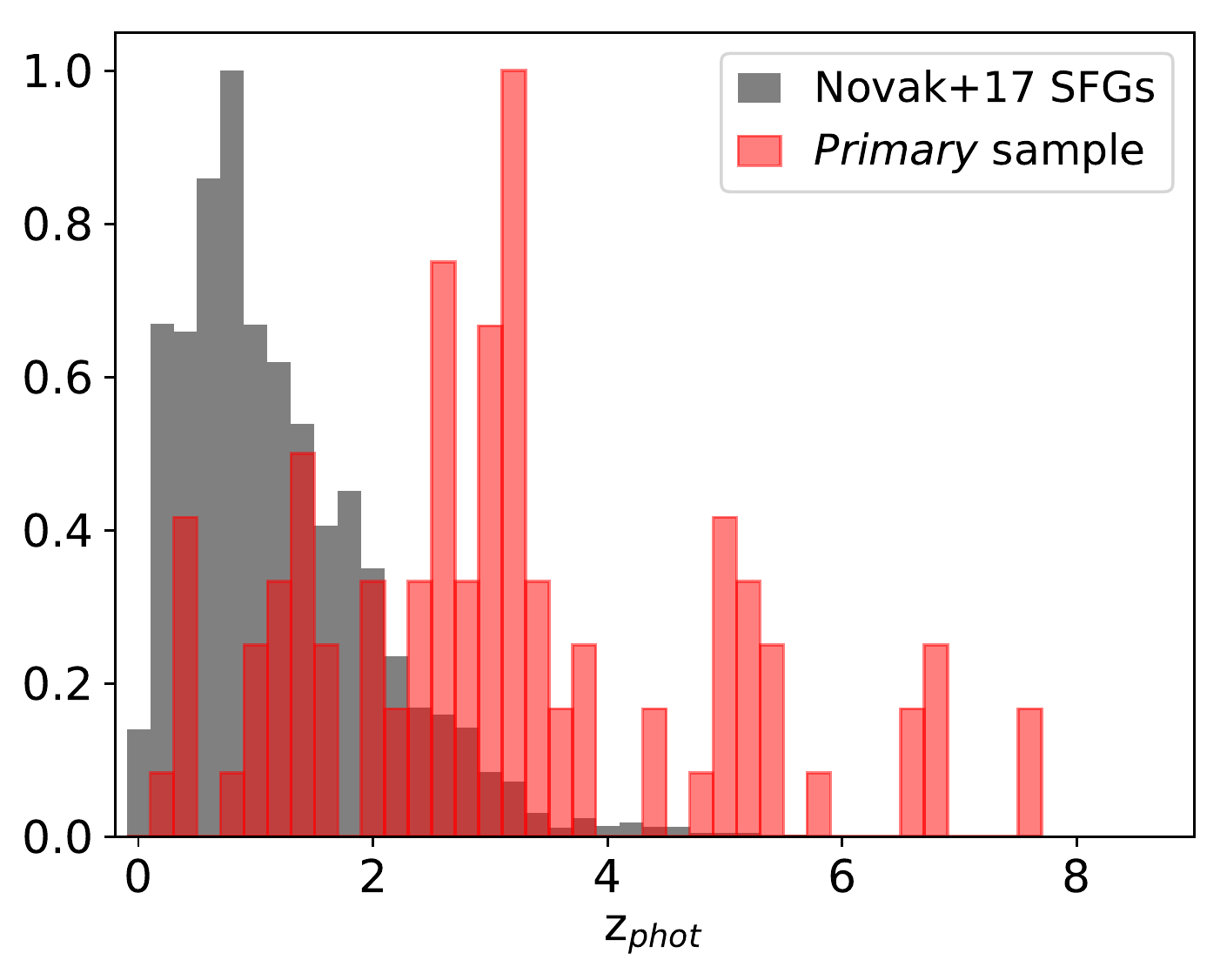}
	\caption{Redshift distributions (normalized to their maximum) of our \emph{primary} sub-sample (red), compared to the complementary sample of radio-selected galaxies with OPT/NIR counterpart \citep{Novak2017} (grey). }
	\label{fig:zhist}
\end{figure}

In Fig.~\ref{fig:zhist} we compare the redshift distribution of our \emph{primary} sources with a complementary radio-selected sample with OPT/NIR counterparts \citep{Novak2017}, taken from the same VLA-COSMOS 3 GHz Large Project. 
The derived redshift distribution and median physical properties (see Table \ref{ext_tab1}) confirm that the bulk of our population of UV-dark radio sources is indeed consisting of dusty SFGs at z$\sim$ 3, but we highlight a tail of 22 newly identified very high-z candidates (z$>$4.5).
We also compared our redshift estimates with those available in the  \citet{Jin2018} catalog and found a mild agreement. 
We stress that the estimates of photometric redshifts in  \citet{Jin2018} were computed using FIR data alone, while in our case we used the entire UV-to-radio SED, including upper limits.

A spectroscopic redshift, from Ly$\alpha$ and sub-mm emission lines, was available in the literature only for one source, COSMOSVLA3$\_$181 (a.k.a AzTEC-C159 \citep{Smolcic2015, GomezGuijarro2018}), and it was consistent with our photometric estimate: z$_{spec}$=4.57 vs. our z$_{phot}$=5.11$\pm$0.4. 
Although this latter comparison is reassuring about the robustness of our photometric reshifts, a spectroscopic follow-up of our entire population of radio-selected UV-dark galaxies would be crucial to the full understanding of their properties.

\begin{figure}[ht!]
	\centering
	\includegraphics[scale=0.35]{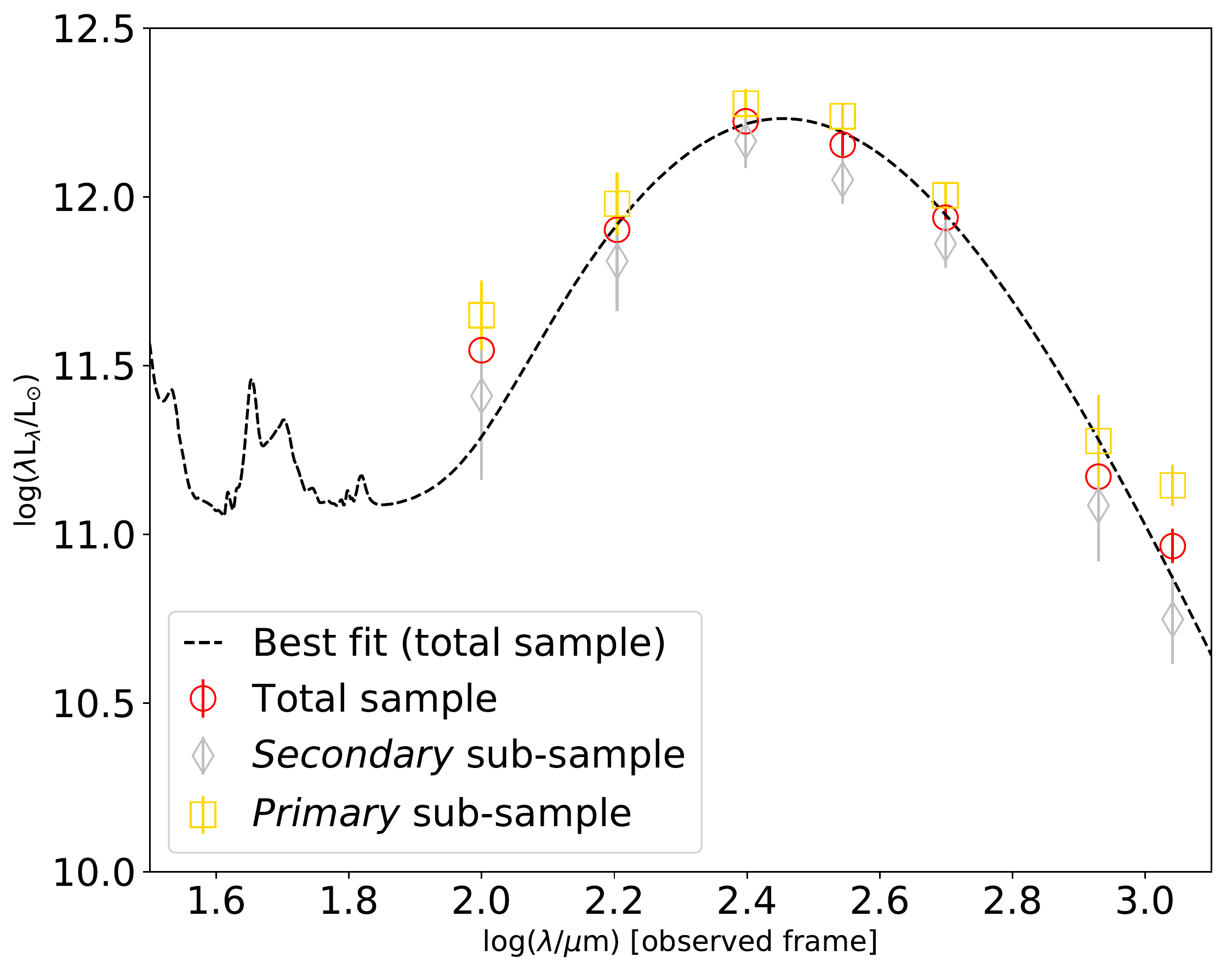}
	\caption{Mean far-infrared SED of the total sample of radio-selected galaxies (red points and dashed curve), the \emph{primary} sub-sample (gold squares), and the \emph{secondary} sub-sample (silver diamonds). 
	}
	\label{fig:stack_herschel}
\end{figure}

Regarding the \emph{secondary} sub-sample, where no photometric information is available at FIR wavelengths for individual galaxies, the constraints on the physical properties necessarily are weaker.
In Fig.~\ref{fig:stack_herschel} we compare the mean stacked FIR SEDs of the three samples (total, \emph{primary}, and \emph{secondary}) cited in this work. 
The points were derived by mean stacking the images of our sources in five Herschel bands (from 100 to 500 $\mu$m), SCUBA 850 $\mu$m and AzTEC 1.1mm, following the procedure by  \citet{bethermin2015}.
We did not apply any correction to account for possible contamination of the stacked flux by clustered neighbors (see Appendix A of \citealp{bethermin2015}), because galaxies in our sample were selected to be isolated (Sec. \ref{sec:sample_sel}).
The almost identical position of the FIR peak in the \emph{primary} and \emph{secondary} stacked samples suggests that the redshift distributions are also likely similar, while the flux in the FIR bands of the \emph{secondary} sample is on average $\sim$35$\%$ lower than the \emph{primary} sample. 
Also, it is reasonable to deduce that their median stellar mass is about 50$\%$ lower, based on the NIR-to-IRAC fluxes (see the gray points in Fig. \ref{fig:medsed}). 
 
A detailed analysis of the physical properties of our sample of UV-dark radio-selected galaxies will be presented in a forthcoming paper (Giulietti et al. in preparation). 

\section{Star-formation rate density} 

\subsection{Correction for incompleteness} \label{sec:correction}

To compute the density of sources and subsequently the SFRD at different cosmic times, we employed the $1/V_{max}$ method \citep{schmidt1968}.
We chose four redshift bins, the highest-redshift one being z$_{phot}>$ 4.5 (see Table \ref{ext_tab2}). 
For each galaxy, we computed the maximum observable volume as:
\begin{equation}
V_{max}=\sum_{z=z_{min}}^{z_{max}} [V(z+\Delta z)-V(z)] \times \frac{C_{A}}{C_{I}}
\end{equation}
where the sum adds together comoving volume spherical shells in small redshift steps of $\Delta z$ = 0.005 between $z_{min}$, which is the lower boundary of the considered redshift bin, and $z_{max}$, which is the minimum between the maximum redshift at which a source would still be included in the sample given the limiting flux of the survey (assumed to be constant over the entire field) and the upper boundary of the redshift bin.
The $C_{A}$ and $C_{I}$ constants take into account the incompleteness due to our selection criteria, and are further described below. 
Finally, to derive the SFRD as a function of cosmic time we added up the $SFR_{IR}$ of all galaxies in a given redshift bin, weighted by their individual $V_{max}$.
SFRD values and their errors, quoted in Table \ref{ext_tab2}, were derived via bootstrap analysis, taking into account the uncertainty on the individual redshifts. In particular, we have first replaced a random number of galaxies in the \emph{primary} sample, we have assigned to each source a redshift by randomly sampling its likelihood distribution with an inverse transform sampling, and finally we have computed the SFRD.
The SFRD values in each redshift bin were defined as the median of the distribution of the values over 400 realizations, while the errors were derived from the 16th and 84th percentiles of the distribution.

The $C_{A}$ parameter is a correction factor that takes into account the observed area:
\begin{equation}
C_{A} = \frac{A_{eff}}{41253 ~deg^{2}}
\end{equation}
where $A_{eff}$=1.38 deg$^{2}$ is the effective unflagged area covered by UltraVISTA observations (see Sec. \ref{sec:sample_sel}) and 41253 deg$^{2}$ is the total sky area.
The $C_{I}$ parameter (correction for incompleteness) is defined as:
\begin{equation}
C_{I} = \frac{(N_{primary}+f\times N_{secondary})}{N_{primary}} \times \frac{N_{tot}}{N_{primary}+N_{secondary}}
\end{equation}
where $N_{primary}$=98 (i.e. the \emph{primary} sub-sample), $N_{secondary}$=99, $N_{tot}$=476 (i.e. the total number of radio sources with no counterpart in the COSMOS2015 catalog), and $f$ is the ratio between the mean SFR$_{IR}$ of the \emph{primary} and \emph{secondary} sub-samples. 
In order to derive the upper and lower boundaries of the SFRD of our UV-dark sample, we made the following considerations about our selection procedure.
Given the purely geometrical nature of our fourth selection step (see Sec. \ref{sec:sample_sel}), i.e. selection of isolated sources in order to ensure uncontaminated photometry, there is no reason to assume that the 197 galaxies analyzed in this work were drawn from a different distribution than the total 476 radio sources with no counterpart in the COSMOS2015 catalog.
On the other hand, the 99 (out of 197) isolated galaxies for which we do not have the individual SED (i.e. the \emph{secondary} sub-sample) might be indeed less star-forming, or at even higher redshift, than the \emph{primary} sub-sample, and might provide a lower contribution to the SFRD. 
We considered three scenarios to derive the $C_{I}$ parameter.

\begin{figure*}[ht!]
	\centering
	\includegraphics[scale=0.9]{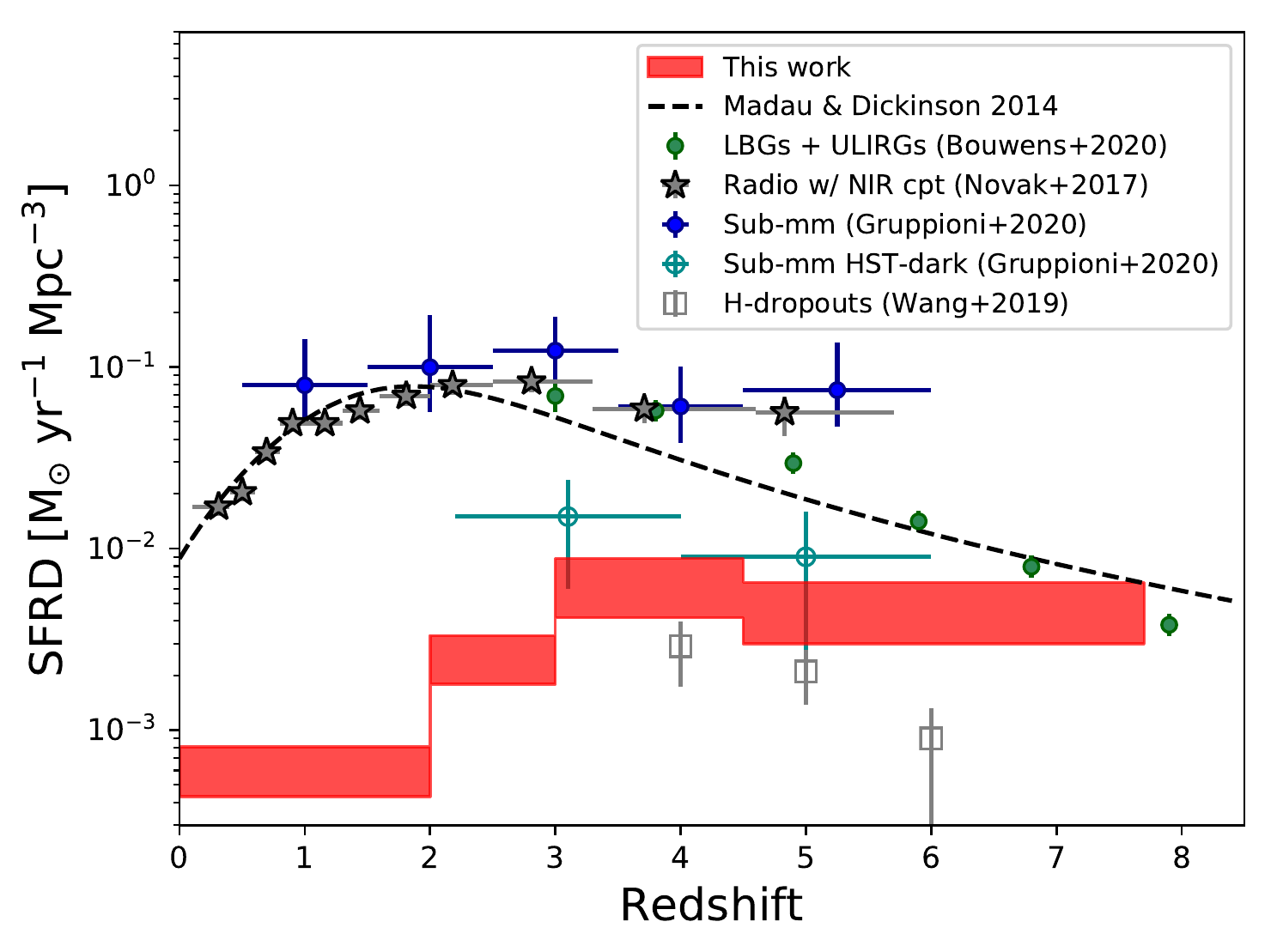}
	\caption{Cosmic star formation rate density (SFRD) history. 
Light-red dashed-shaded regions indicate our confidence interval, depending on the assumptions on the incompleteness correction (see Methods).
The black dashed line indicates the estimate by  \citet{madau2014}, scaled to a  \citet{chabrier2003} IMF. 
At z$>$3 the SFRD estimate is mainly based on LBGs \citep{bouwens2012}.
As a comparison, we also show the samples of: H-dropouts by \citet{wang2019}, radio-selected galaxies with optical counterpart by \citet{Novak2017},  serendipitous ALMA Band-7 (860 or 1000 $\mu$m) continuum detections from ALPINE \citep{gruppioni2020}, LBGs in the ASPECS field (including contribution from ULIRG-type galaxies, \citealp{bouwens2020}), and HST-dark galaxies from ALPINE \citep{gruppioni2020}.
	}
	\label{fig:sfrd}
\end{figure*}

\begin{itemize}
\item \emph{Case 1}: the \emph{primary} sub-sample is fairly representative of the entire radio-selected UV-dark population. Under this assumption, $f=1$ in Eq. (3).
The SFRD values corresponding to this scenario are reported in Table \ref{ext_tab2} and are plotted (added to their uncertainties) as the upper boundary in Fig. \ref{fig:sfrd}
\item \emph{Case 2}: the galaxies in the \emph{secondary} sub-sample do not contribute at all to the SFRD. Under this assumption, $f=0$ in Eq. (3). 
As demonstrated in the previous section, this extreme scenario is not realistic, since we did detect signal in Herschel stacked images of the \emph{secondary} sub-sample. 
Therefore we discarded this hypothesis and we did not report it at all in Fig. \ref{fig:sfrd}.
\item \emph{Case 3}: we assume that the redshift distributions of the \emph{primary} and  \emph{secondary} sub-samples are similar, as hinted by the relative positions of the FIR peaks (see Sec. \ref{sec:individual_SEDs}), and we fix \emph{f} to the ratio between the stacked FIR fluxes of the two sub-samples (Fig. \ref{fig:stack_herschel}).
In this scenario $f=0.65$ in Eq. (3) and the resulting SFRDs are a factor $\sim$1.2 lower than the upper limit set by the \emph{case 1}.
The SFRD values corresponding to the \emph{case 3} scenario, minus their uncertainties, are plotted as the lower boundaries in Fig. \ref{fig:sfrd}.
\end{itemize}

\begin{table}
	\caption{SFRD$^{a}$}
	\centering
    	\label{ext_tab2}
	\begin{center}
    	\begin{tabular}{l l}
    	\hline \hline
	z$_{phot}$ bin          & SFRD [M$_{\odot}$ yr$^{-1}$ Mpc$^{-3}$] \\ 
	\hline
        $[0.0 - 2.0]$		& (6.8 $\pm$ 1.3)$\times 10^{-4}$\\
        $[2.0 - 3.0]$		& (2.8 $\pm$ 0.5)$\times$10$^{-3}$\\
        $[3.0 - 4.5]$		& (7.1 $\pm$ 1.7)$\times 10^{-3}$\\
        $[>4.5]$		& (5.2 $\pm$ 1.3)$\times 10^{-3}$\\	
    	\hline \hline
    	\end{tabular}
	\begin{tablenotes}
	$^{a}$ The tabulated values, added to their uncertainties, correspond to the upper boundary in Fig. \ref{fig:sfrd}. The SFRDs plotted as lower boundary are a factor $\sim$1.2 lower.
	\end{tablenotes}
	\end{center}
\end{table}

\subsection{Results} \label{sec:sfr}

After accounting for our selection function, we estimated the number density \emph{n} of all UV-dark radio sources at $z>3$ to be (1.3$\pm$0.3)$\times$10$^{-5}$ Mpc$^{-3}$.
If we only consider the highest redshift bin ($z>4.5$), \emph{n} = (0.5$\pm$0.1)$\times$10$^{-5}$ Mpc$^{-3}$, which is comparable to other estimates in the literature derived from smaller samples of dark galaxies at z$>$ 5 \citep{riechers2020, gruppioni2020}.
Existing semi-analytical models \citep{henriques2015} and hydrodynamical simulations \citep{snyder2017, pillepich2018} in the literature do not predict the early formation of such a large number of massive, dusty galaxies, and underestimate their number density by one to two orders of magnitude \citep[see also][]{wang2019}, with respect to our findings.
Interestingly, we notice that the number density of our UV-dark radio-selected galaxies at high-redshift is comparable to that of massive quiescent galaxies at 3$<$z$<$4 ($\sim$2$\times$10$^{-5}$ Mpc$^{-3}$) \citep{straatman2014, schreiber2018b}.
UV-selected galaxies at z $>4$ are presumed not to be abundant and star-forming enough to produce the earliest known massive quiescent galaxies \citep{straatman2014}, suggesting that most of the star formation in the progenitors of quiescent galaxies at high-z was obscured by dust.
However, it is difficult to establish a conclusive connection between high-z quiescent galaxies and the various samples of dusty starburst galaxies identified up to z $\sim$6 \citep{Marrone2018} (e.g. the so-called 500 $\mu m$ or 850 $\mu m$ risers: \citealp{Cox2011, Riechers2013, Ivison2016, Riechers2017, Ma2019}), because of their low and uncertain number density, which is one order of magnitude lower than that of our UV-dark sources.
The number density and high levels of star-formation of our sample of UV-dark sources could imply an evolutionary link between this population and that of high-z quiescent galaxies, where a significant fraction of the latter ones could be the descendants of the former ones at higher redshifts.

\citet{wang2019} reported,  for their sample of H-droupout-selected dusty galaxies, a space density ($\sim$2$\times$10$^{-5}$ Mpc$^{-3}$ at $z>3$) similar to that of our high-redshift UV-dark galaxies, in the same redshift range. 
However, comparing the two populations, we find that our radio-selected galaxies produce stars at a rate which is on average three times higher.
In fact, the median L$_{IR}$ of our galaxies at z$>$3 is (6.3$\pm$0.5)$\times$ 10$^{12}$, which is about three times higher than the median infra-red luminosity of (2.2$\pm$0.4)$\times$ 10$^{12}$ reported by Wang et al. for their sample.
This means that their contribution to the cosmic SFRD is also higher. 
In particular, it reaches 7.0$\times$ 10$^{-3}$ M$_{\odot}$ yr$^{-1}$ Mpc$^{-3}$ in the redshift range $3.0<z<4.5$, and 5.4$\times$ 10$^{-3}$ M$_{\odot}$ yr$^{-1}$ Mpc$^{-3}$ at $4.5<z<7.7$, with uncertainties of the order of 20$\%$ (Fig.~\ref{fig:sfrd} and Table \ref{ext_tab2}), therefore doubling the contribution of the population of H-dropouts previously cited at the same cosmic epoch \citep{wang2019}.
A similar result was reported for a sample of HST-dark galaxies identified at sub-mm wavelengths in the ALPINE fields \citet{gruppioni2020}.

We compared our own selection criteria also with those by \citet{wang2019}, again to estimate the possible overlap between the two samples (see the Appendix), and we concluded that at least $\sim$82$\%$ of the UV-dark radio-selected galaxies are not consistent with the H-dropout selection by \citet{wang2019} and therefore constitute a different galaxy population.
It is worth noting that we find a similar percentage (83$\%$) when focusing only on the common redshift range between the two samples (z$>$3).

We compared the trend toward high redshift of the cosmic SFRD of our radio-selected UV-dark galaxies with the estimate based on galaxies identified in the rest-frame UV regime \citep{madau2014, bouwens2020}\footnote{\citet{bouwens2020} include the SFRD contribution from ULIRG-type galaxies in the ASPECS volume, derived from various sub-mm surveys.}. 
At $3.0<z<4.5$, the contribution of UV-dark radio galaxies to the SFRD corresponds to $\sim$10-25$\%$ of the SFRD of UV-bright galaxies.
This fraction rises to $\sim$25-40$\%$ at z$>4.5$, where we identified 22 very high-z candidates, whose SEDs are illustrated in Fig.~\ref{fig:highz} along with their redshift likelihood distributions.
We stress that the actual contribution of radio-selected UV-dark galaxies to the cosmic SFRD could be even higher, because our estimates are derived without any extrapolation to lower fluxes (hence SFRs) than our selection, which would have been extremely uncertain.
In fact, the correction for incompleteness which we apply only takes into account how representative the \emph{primary} sample is with respect to our total sample of radio-selected UV-dark galaxies.

The cosmic SFRD of UV-dark radio-selected galaxies is flatter than that of UV-bright galaxies and similar to the complementary sample of radio-selected galaxies \citep{Novak2017} with OPT/NIR counterparts and to that of serendipitously detected sub-mm galaxies in the ALPINE fields from \citet{gruppioni2020} \citep[see also][]{loiacono2020}.
The estimates of the SFRD based on far-IR/sub-mm and radio data, from different works \citep[see also][]{rowanrobinson2016}, show an almost flat trend at z$>$2, suggesting a significant contribution of dust-obscured activity which cannot be recovered by the dust-extinction corrected UV data.
Our results quantify which fraction of the missing amount of star-formation activity could be explained by UV-dark galaxies. 

\section{Summary}

Our findings have several implications. 
We demonstrated that starting from deep radio surveys, it is possible to identify a population of massive (median M$_{star} \sim$ 1.7 $\times 10^{11}$ M$_{\odot}$), extremely dust-obscured (A$_{V} \sim$4) SFGs at $z\sim 3$, which are invisible in current optical surveys and near the detection limit of the deepest available NIR surveys.
The radio selection turned out to be particularly effective in identifying candidates at very high redshift (z$>$4.5), whose number density is under-predicted by current simulations and which provide a significant contribution to the cosmic SFRD. 
The comparison of different selections of UV-/HST-dark galaxies from the literature showed only a partial overlap between the various samples, suggesting a possible diversity of galaxy populations under the common \emph{dark} label and the need of a multi-wavelength approach to the search of such objects.
These results highlight the limits of our current understanding of the processes that govern galaxy formation.
To enhance our knowledge of the dust-obscured part of the high-redshift Universe, observational efforts, especially spectroscopic follow-ups with current and upcoming facilities like ALMA and the James Webb Space Telescope,  should be focused on UV-dark radio-selected galaxies, since the confirmation of their redshifts, along with information on chemical abundances, stellar masses and dust properties would prove essential to our understanding of massive galaxies assembly.\\

\begin{figure*}[!ht]
	\centering
	\includegraphics[scale=0.3]{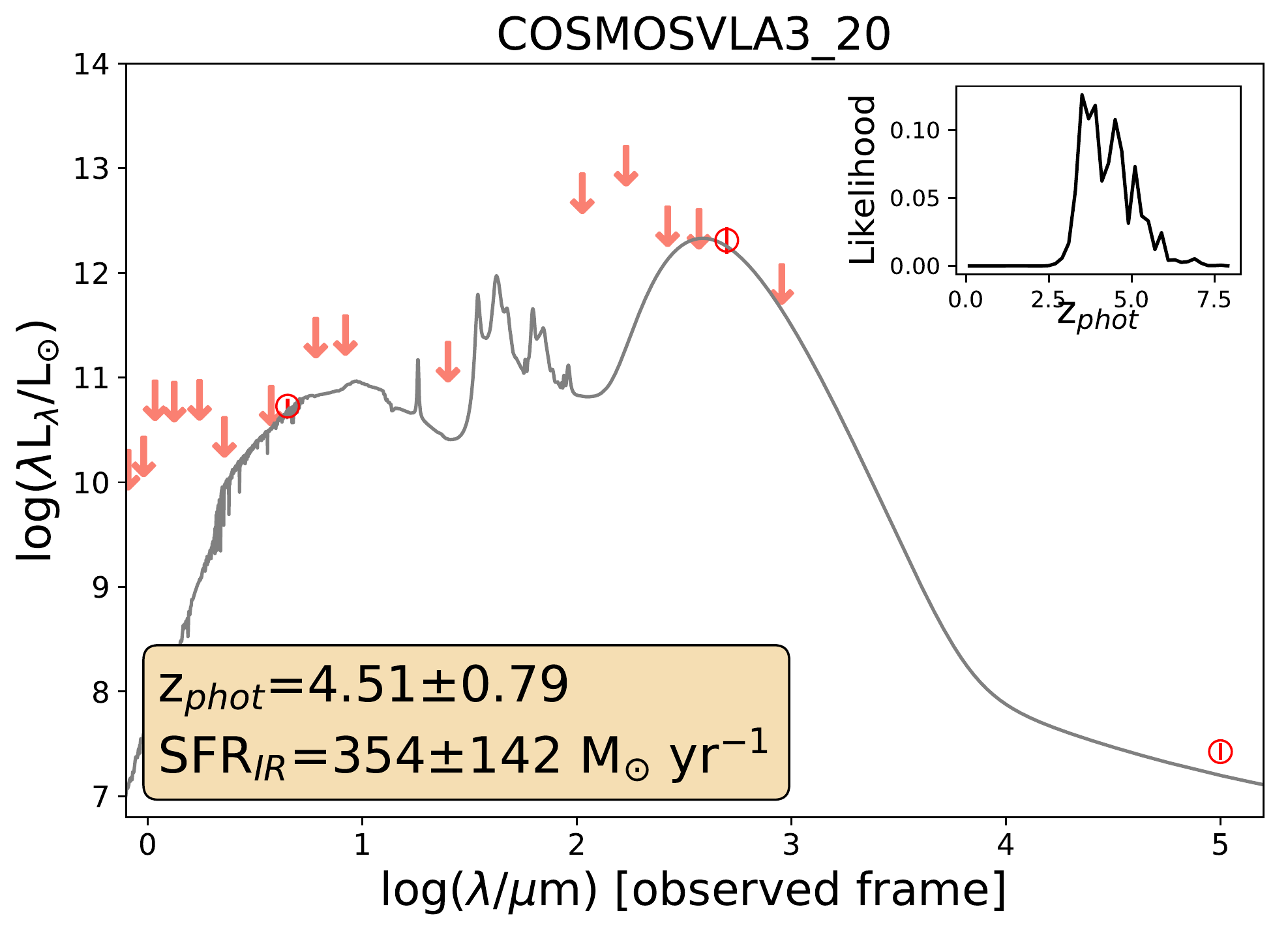}\includegraphics[scale=0.3]{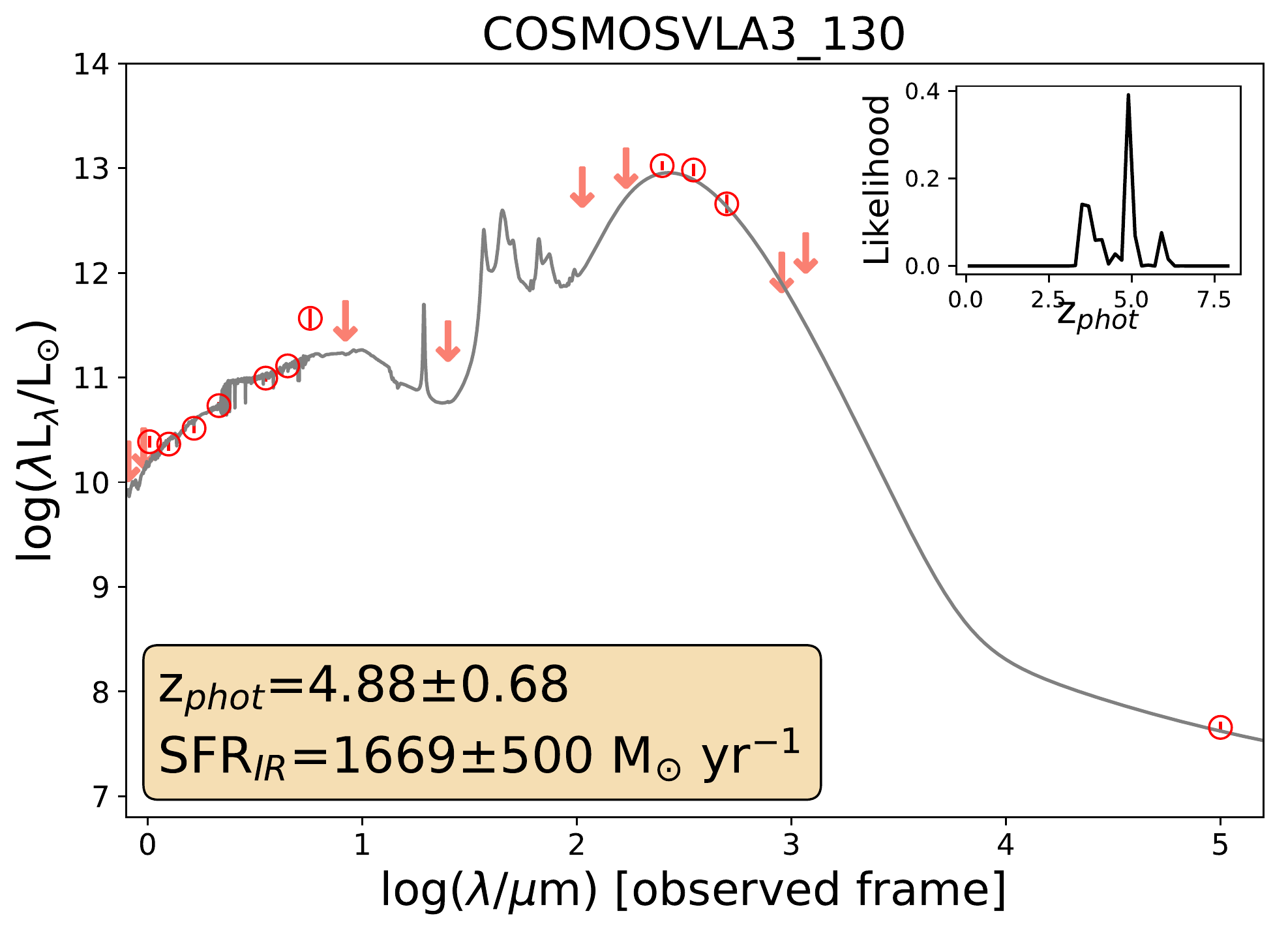}\includegraphics[scale=0.3]{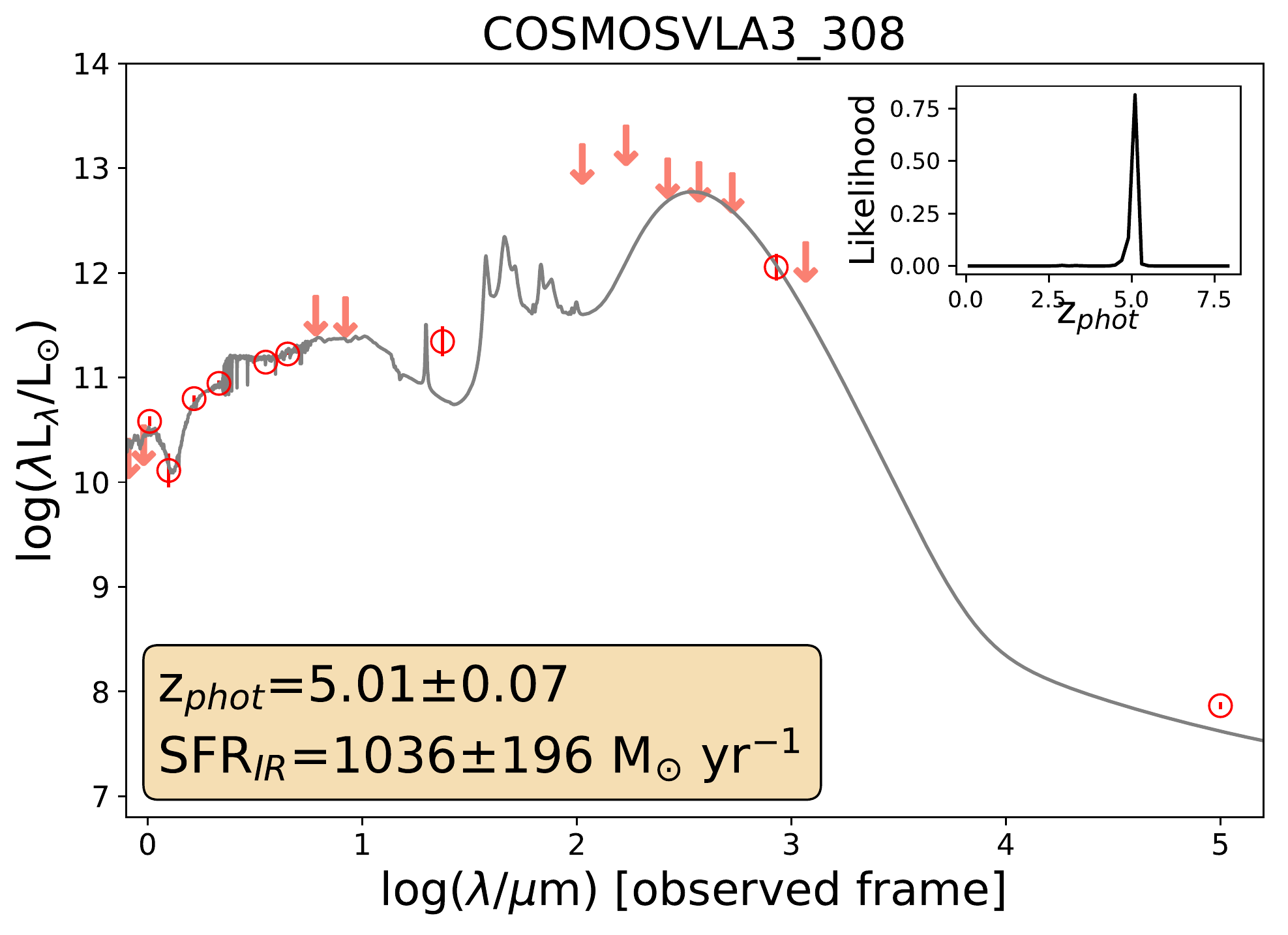}
	\includegraphics[scale=0.3]{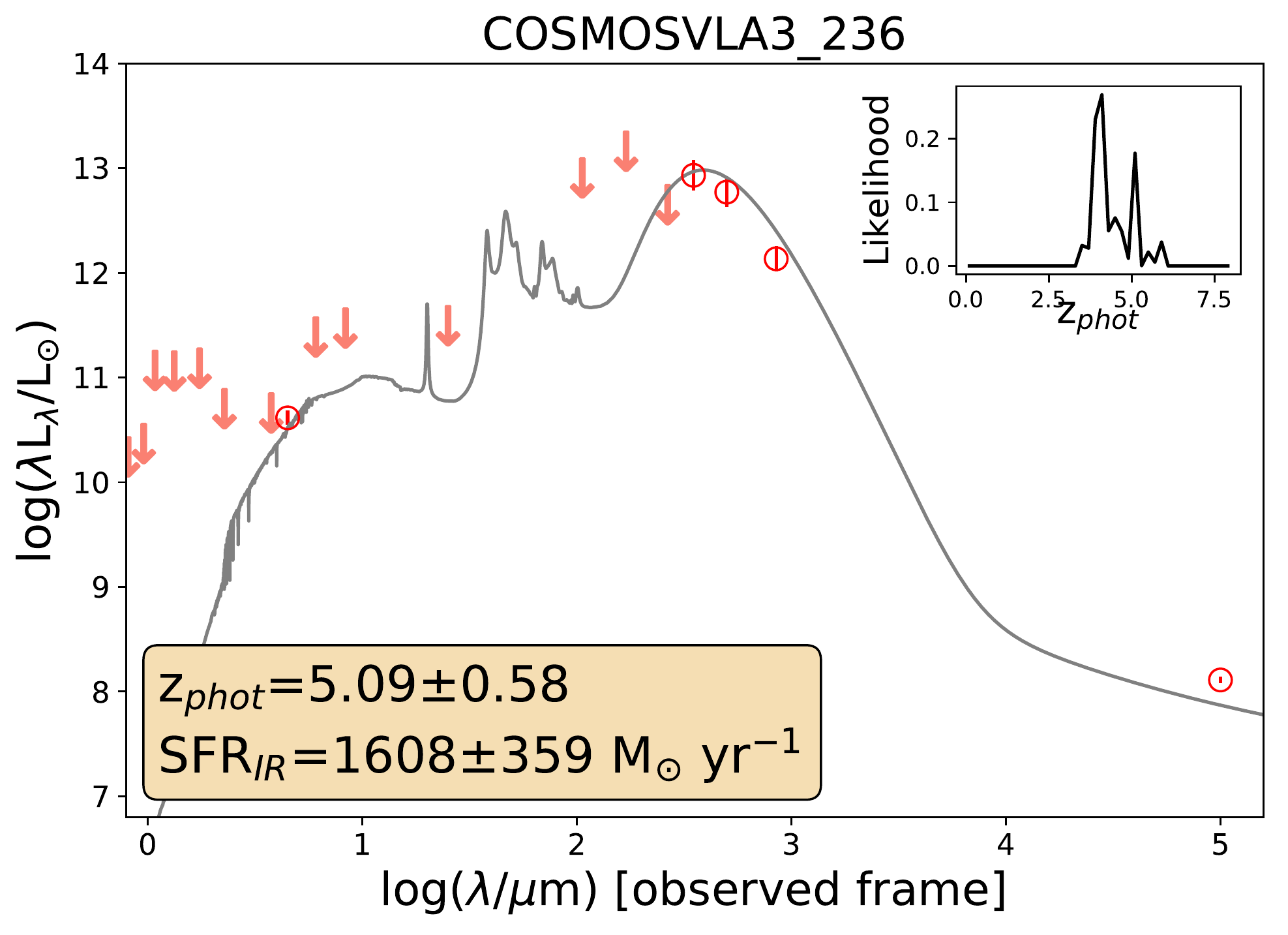}\includegraphics[scale=0.3]{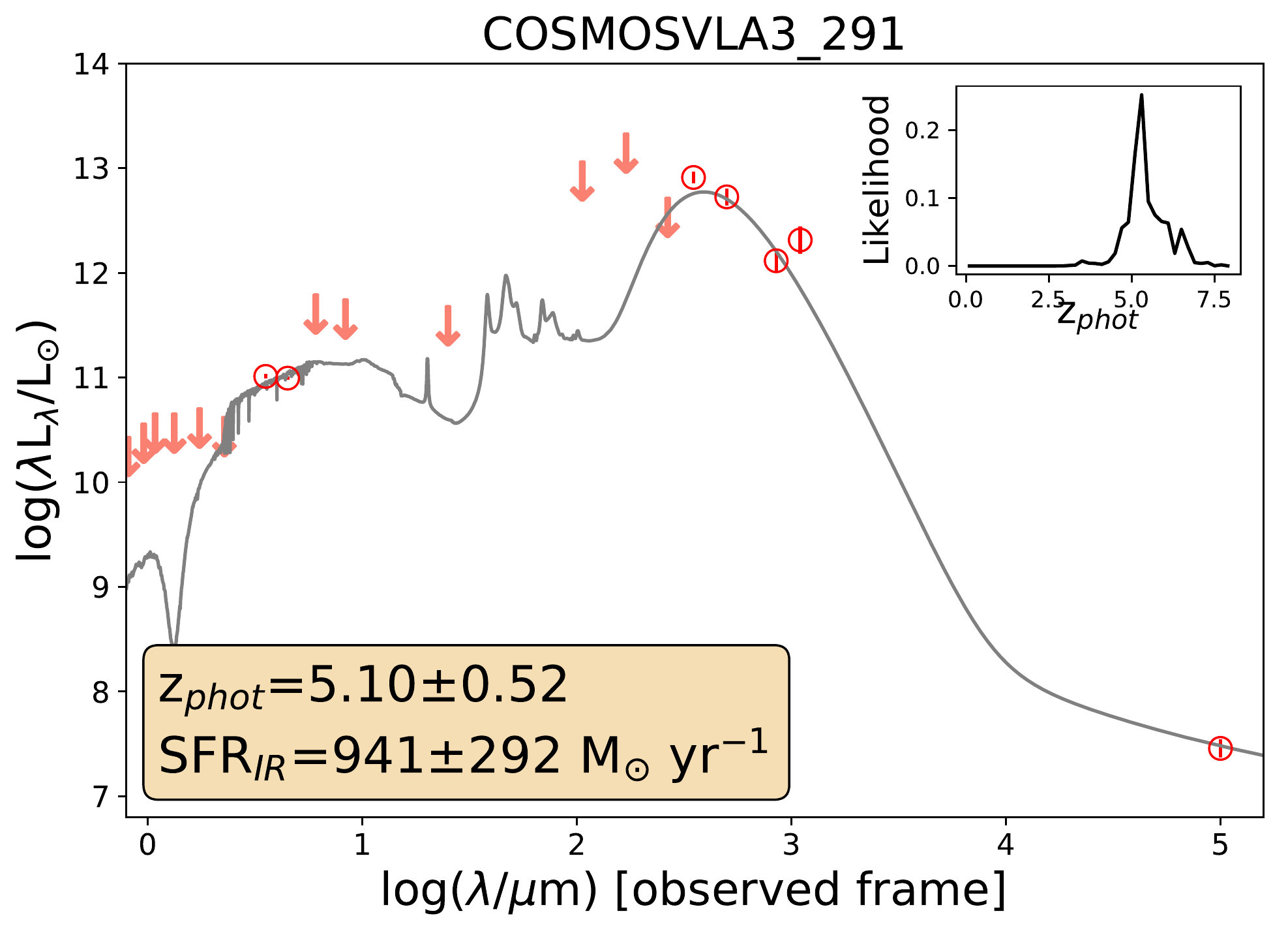}\includegraphics[scale=0.3]{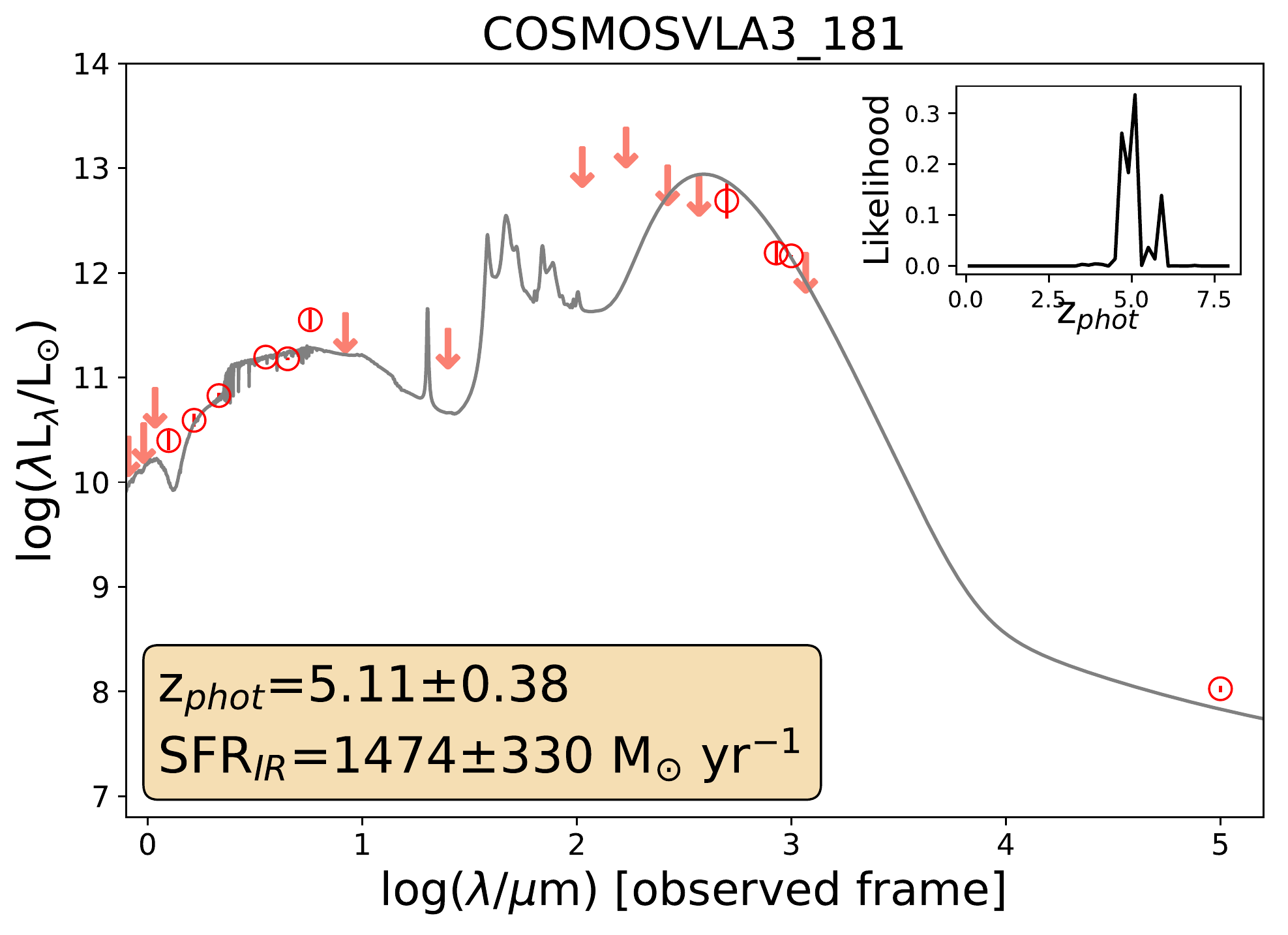}
	\includegraphics[scale=0.3]{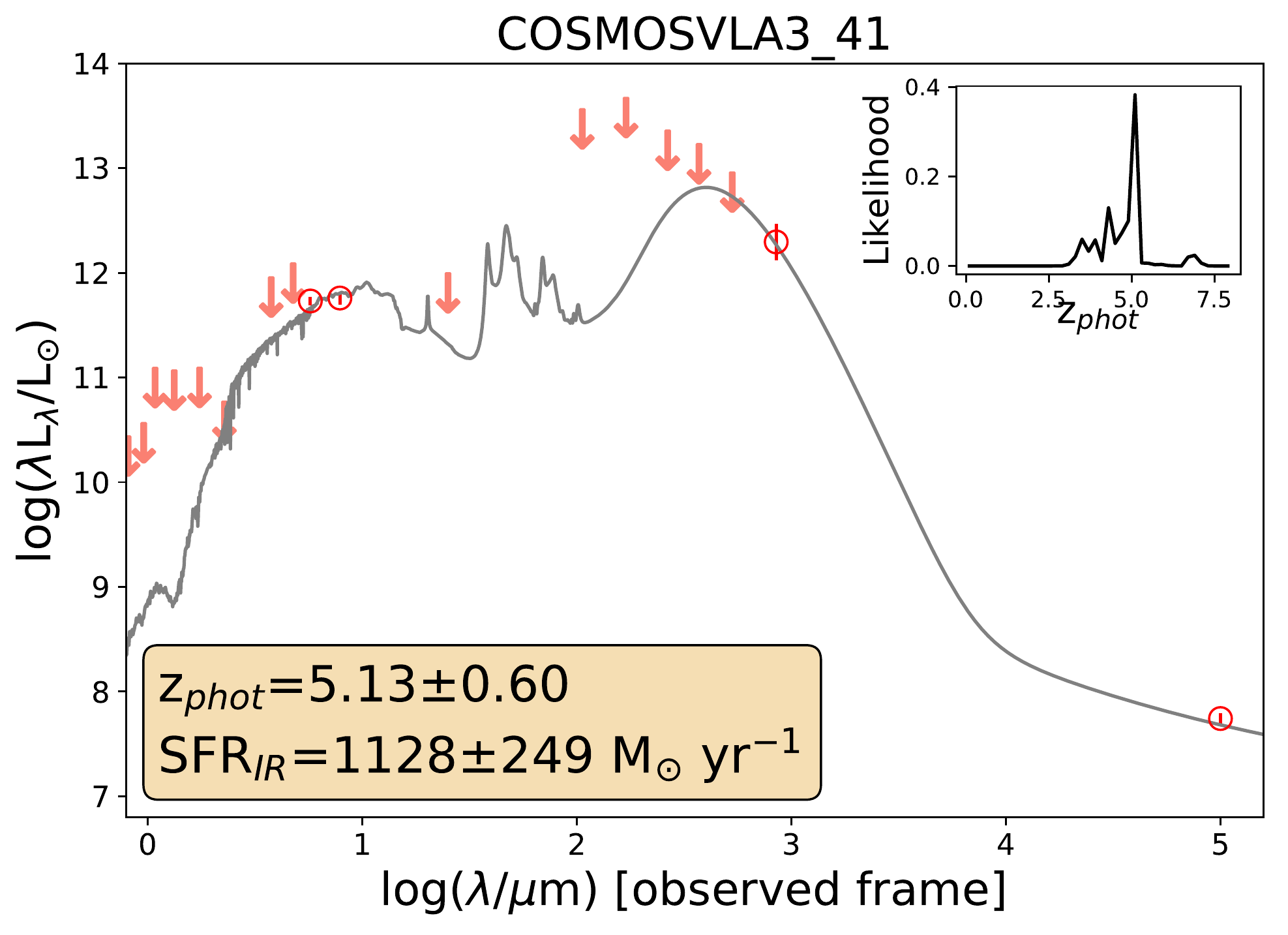}\includegraphics[scale=0.3]{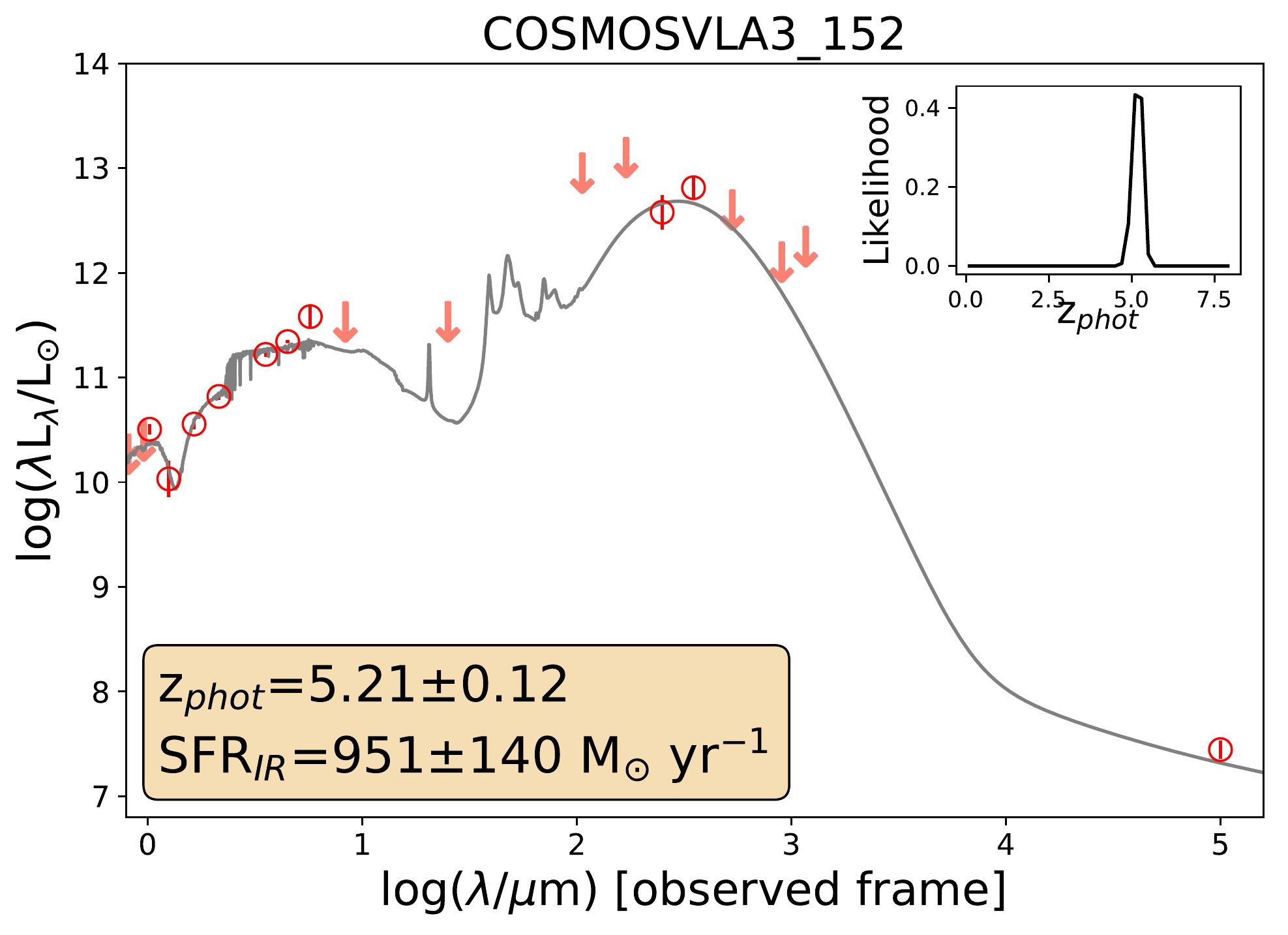}\includegraphics[scale=0.3]{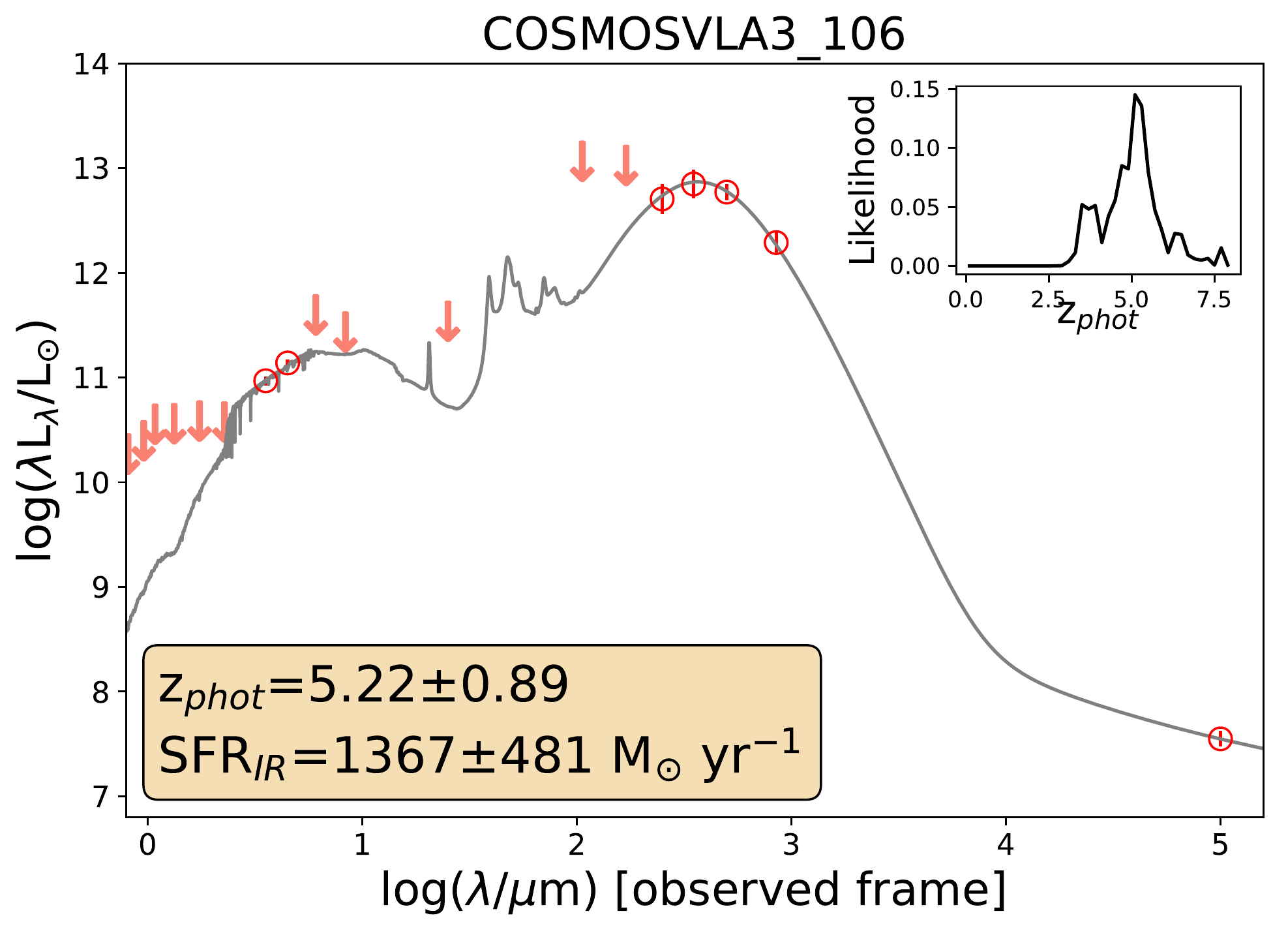}
	\includegraphics[scale=0.3]{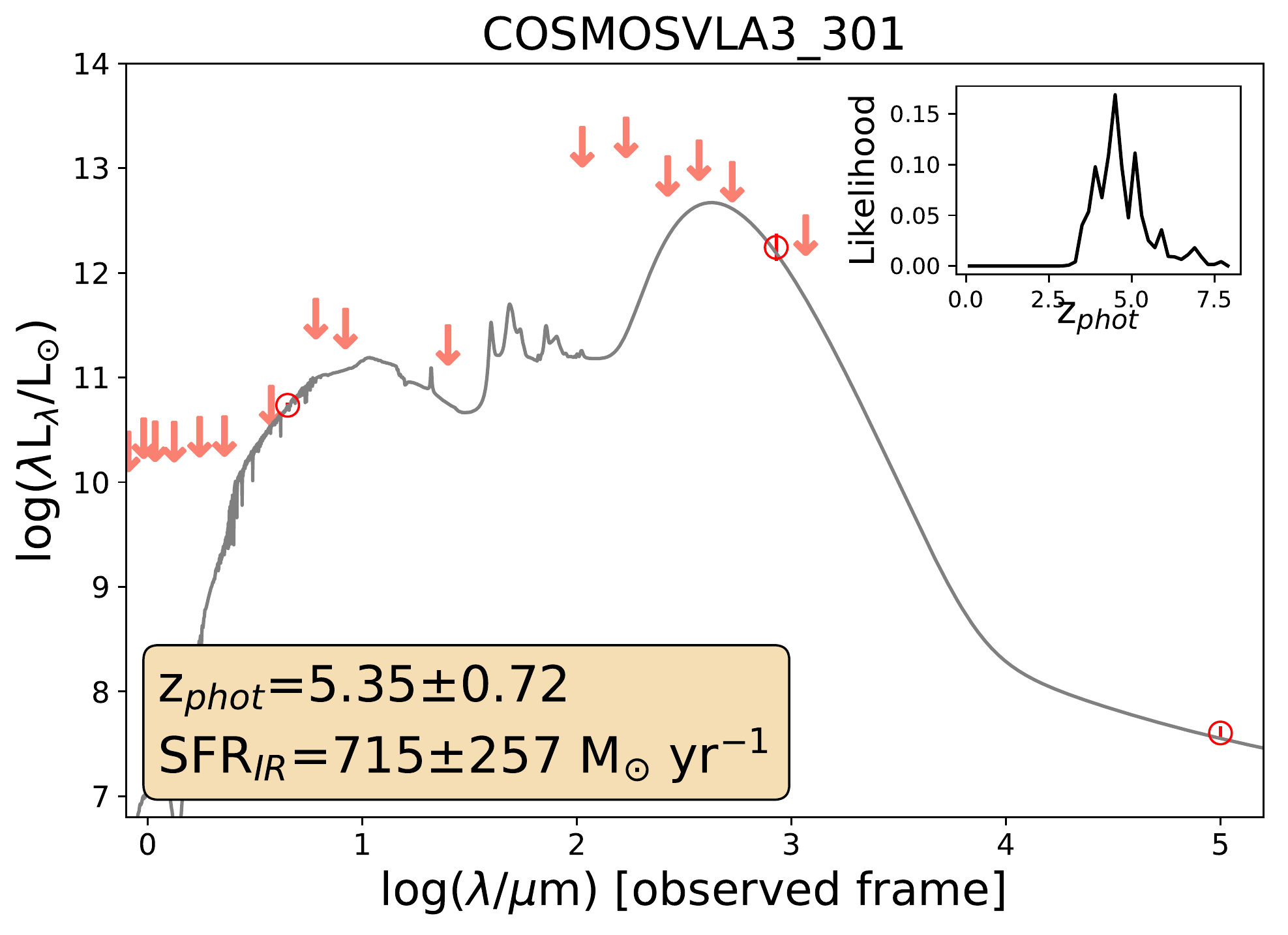}\includegraphics[scale=0.3]{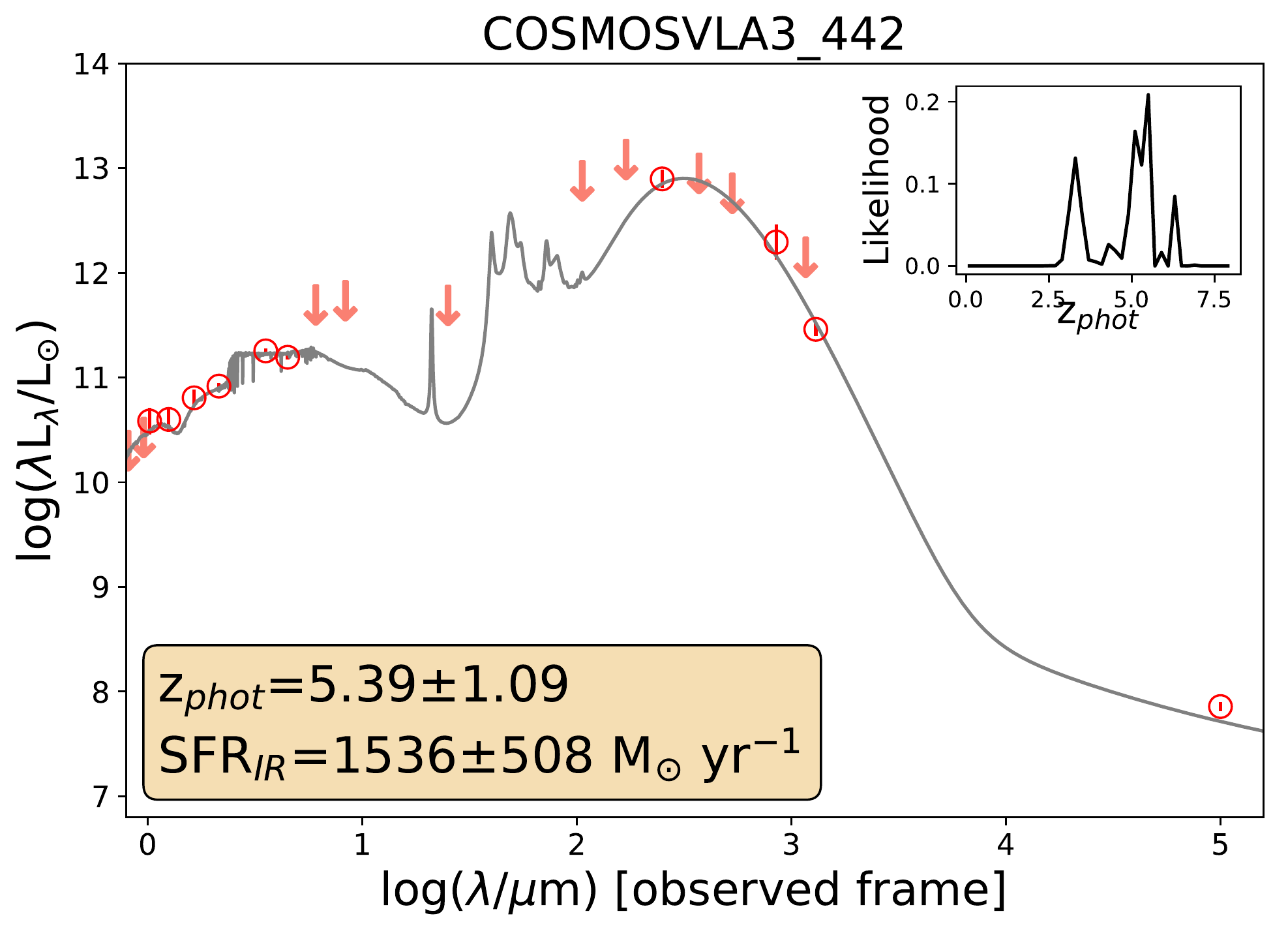}\includegraphics[scale=0.3]{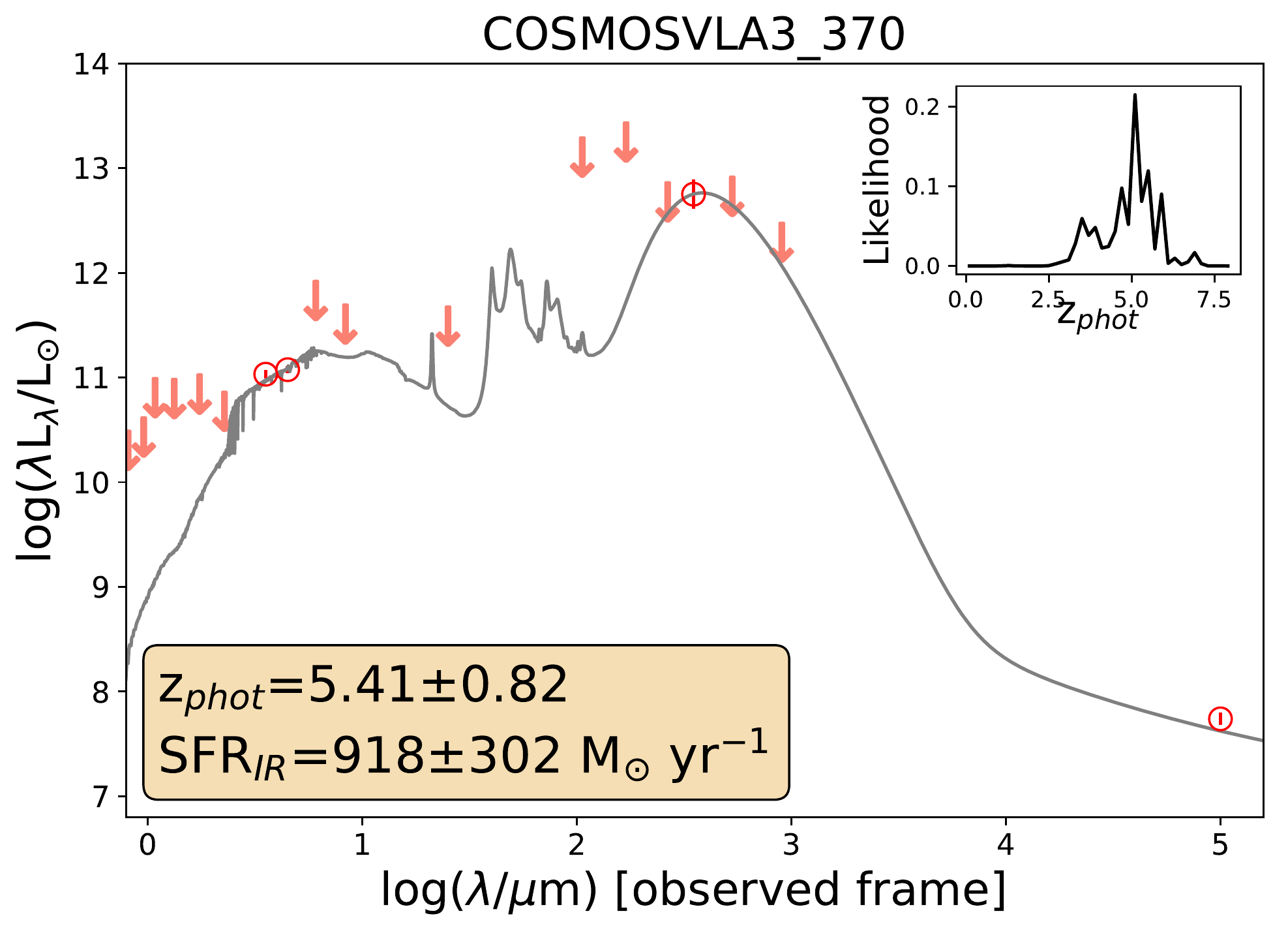}
	\includegraphics[scale=0.3]{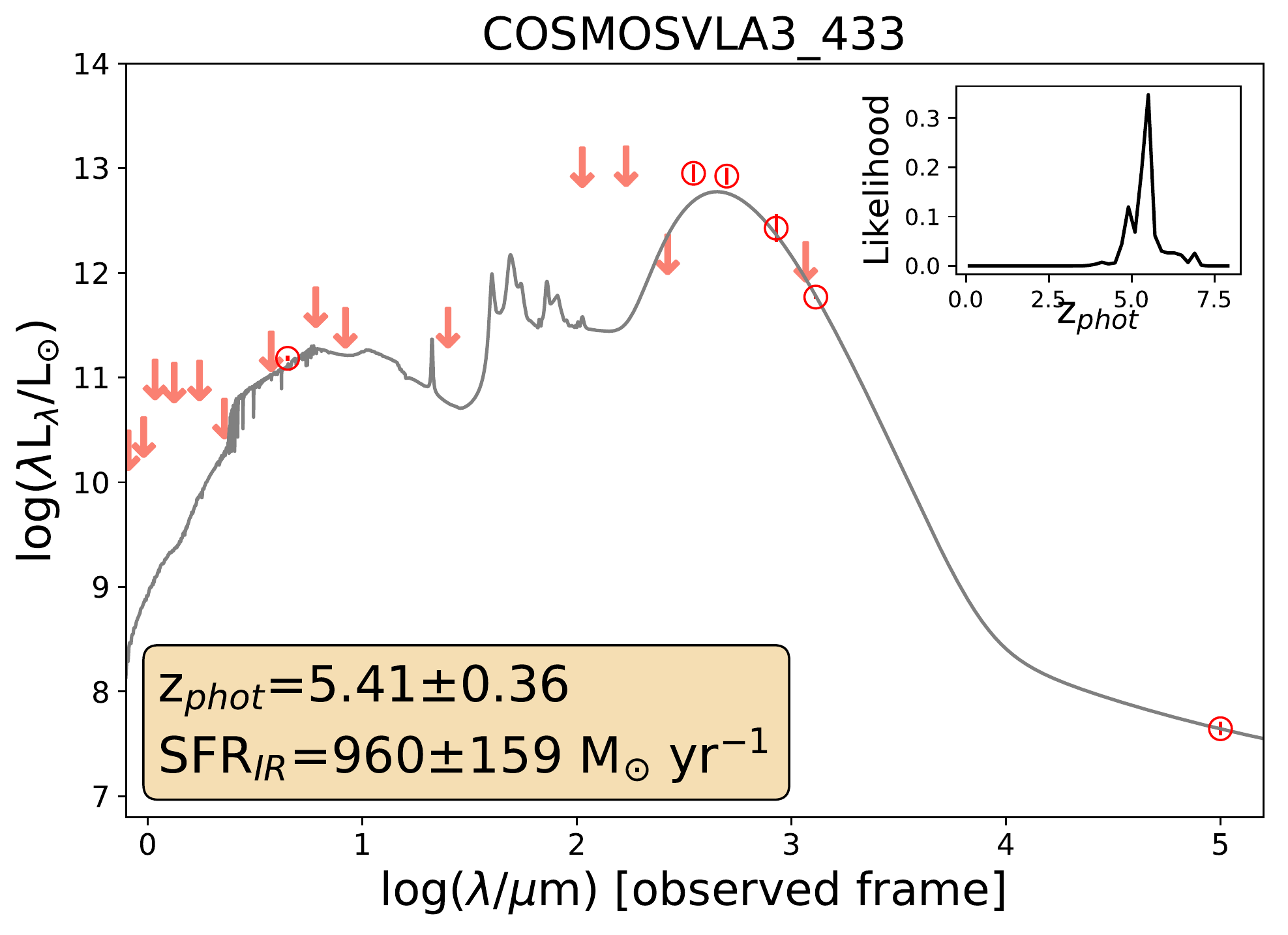}\includegraphics[scale=0.3]{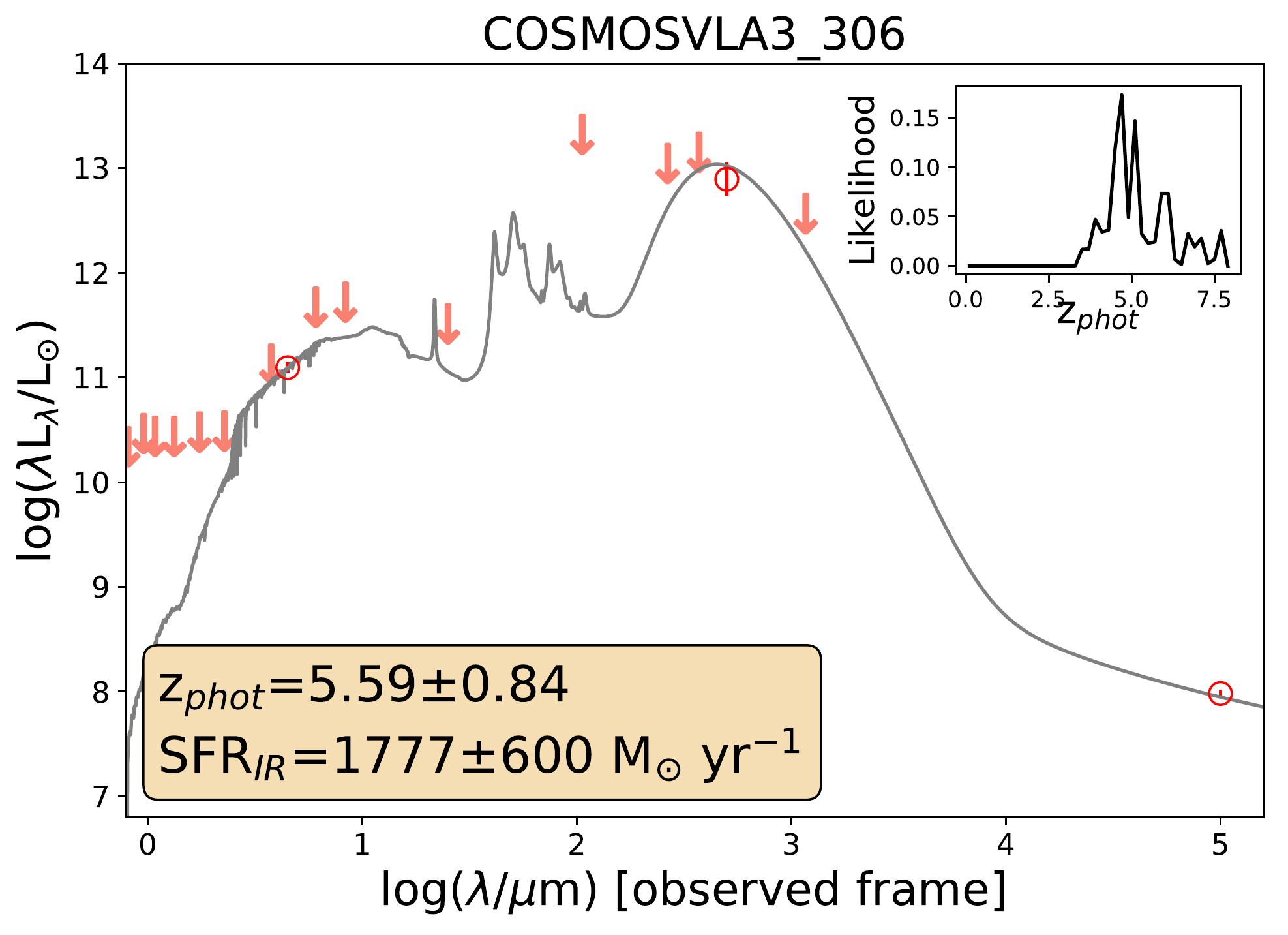}\includegraphics[scale=0.3]{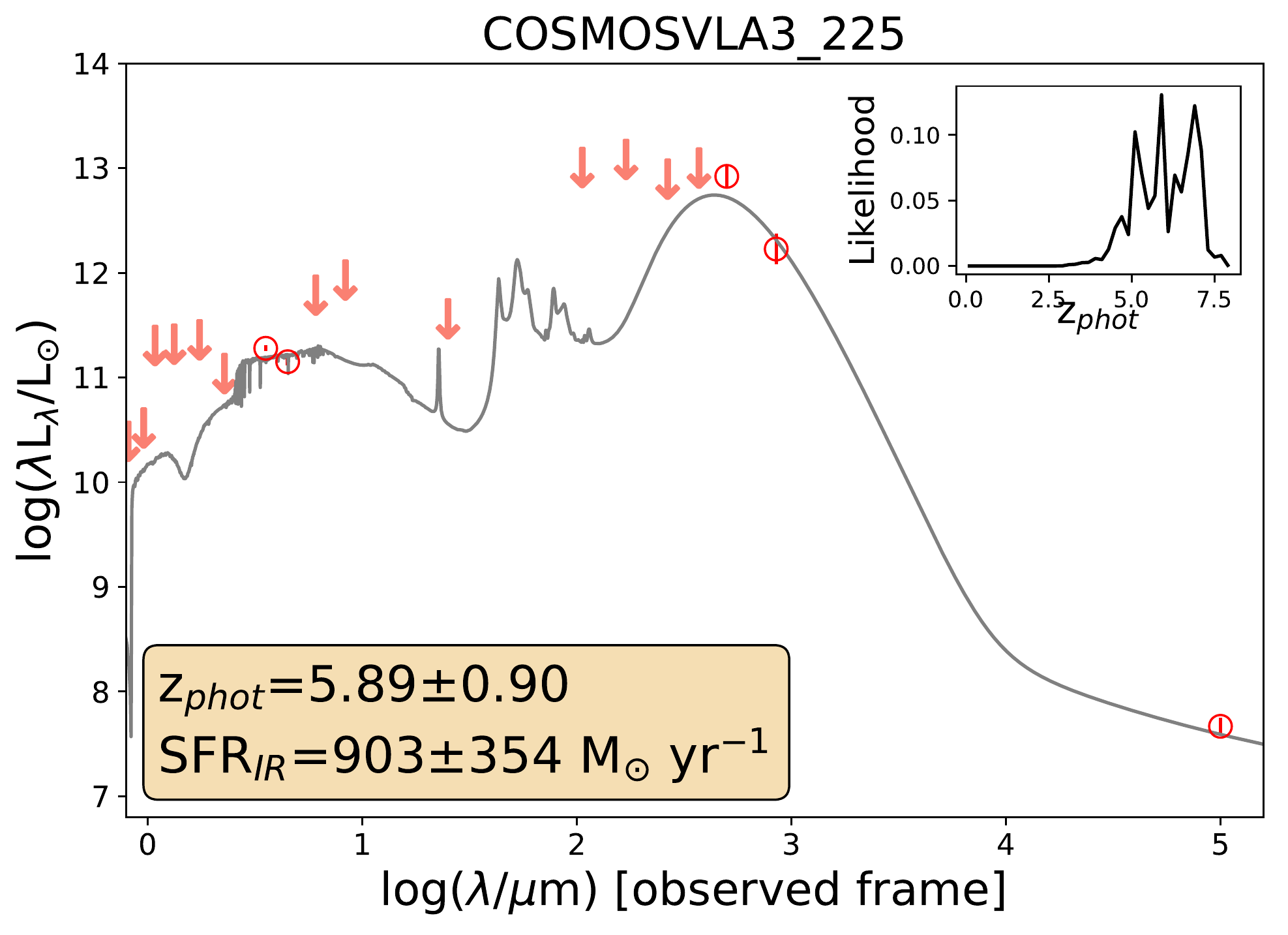}
	\caption{Observed (points as in Fig. \ref{fig:medsed}) and best-fit (black line) SEDs of the highest-redshift galaxies from our \emph{primary} sub-sample (z$\geq$4.5). Error bars are plotted for each data-point, although in some cases they are smaller than the points. For each galaxy we quote the SFR$_{IR}$ and show the likelihood distribution of the photometric redshift in the inset (continued in the next page).}
	\label{fig:highz}
\end{figure*}

\begin{figure*}[!ht]
	\includegraphics[scale=0.3]{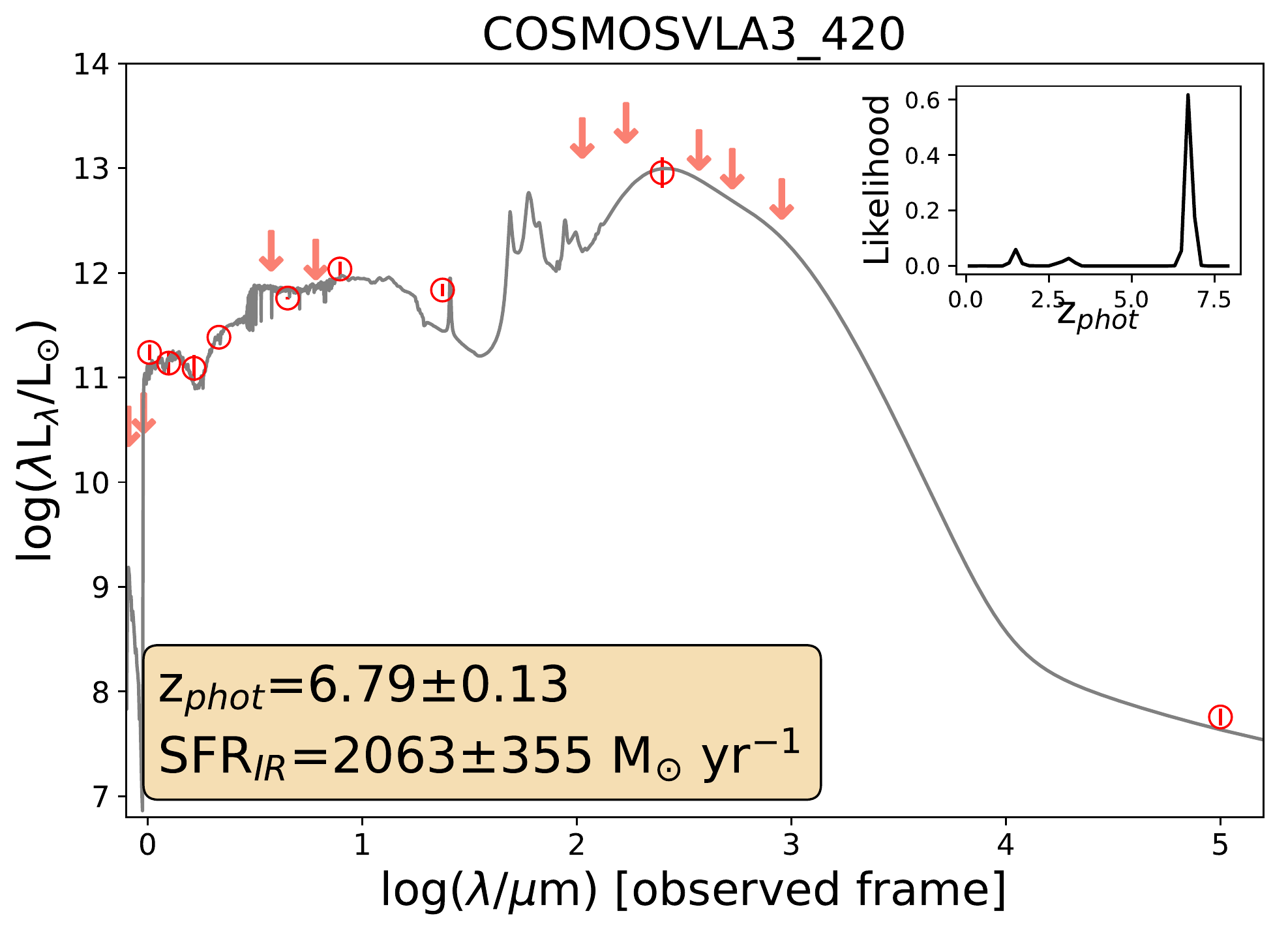}\includegraphics[scale=0.3]{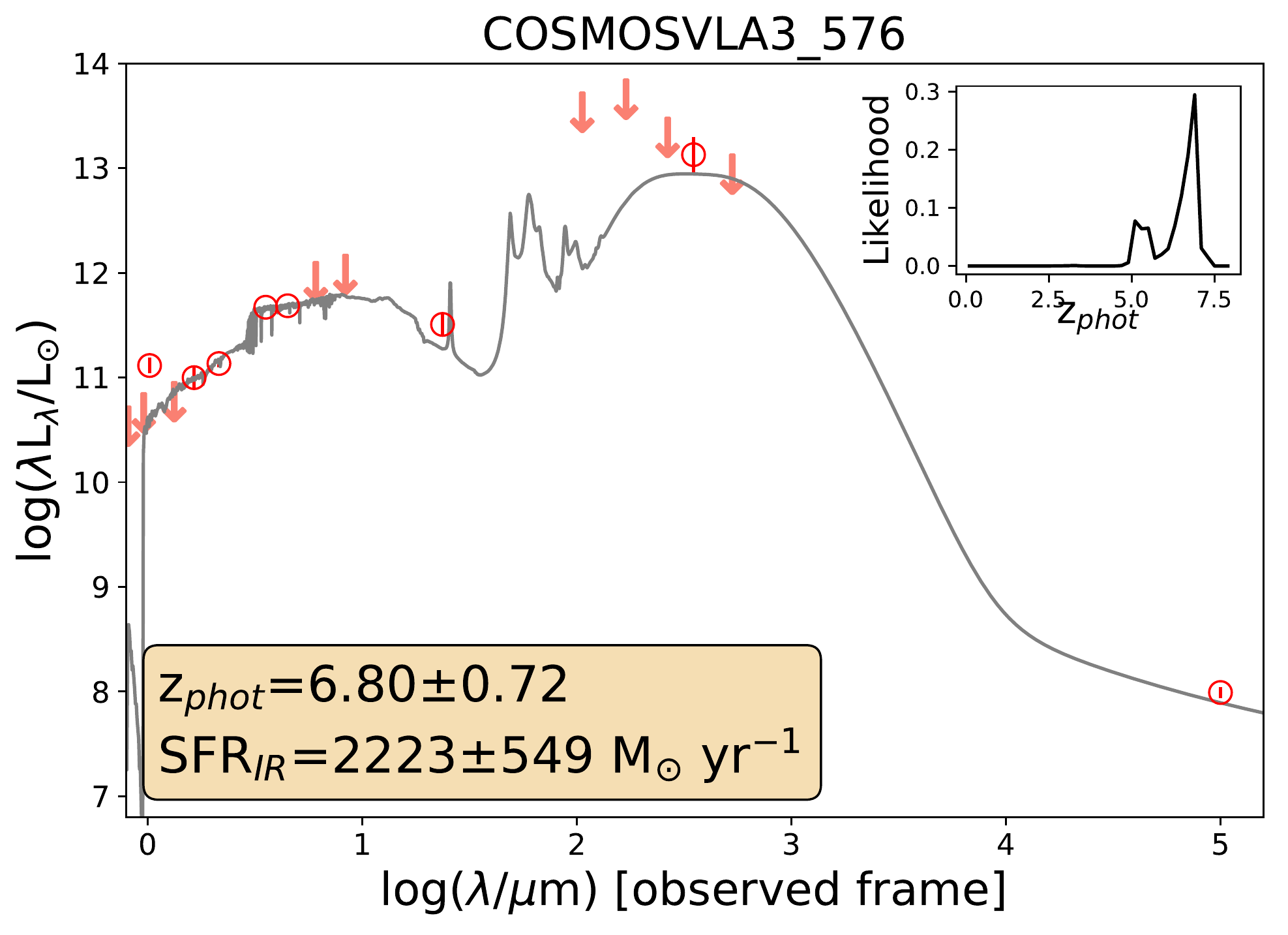}\includegraphics[scale=0.3]{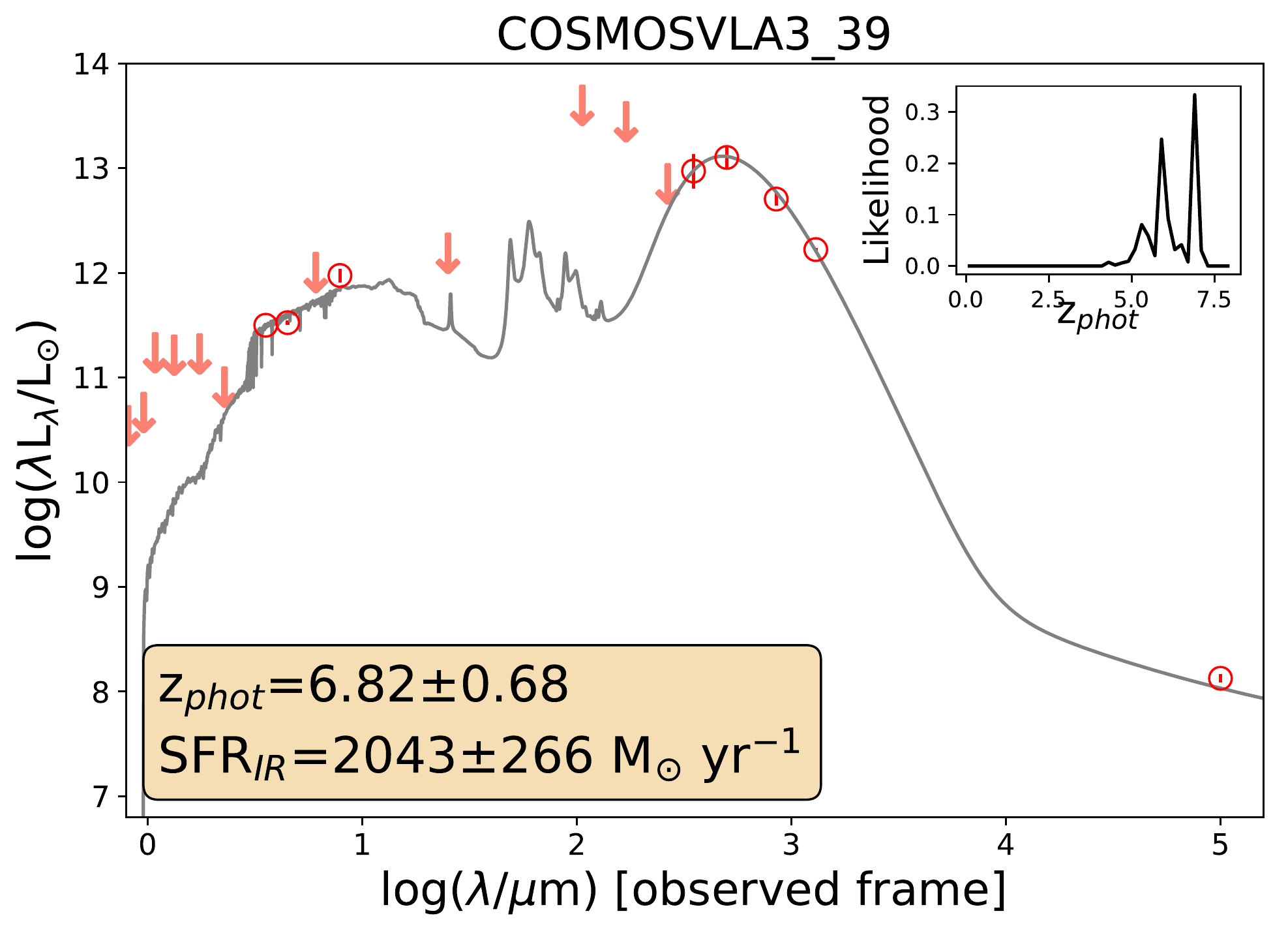}
	\includegraphics[scale=0.3]{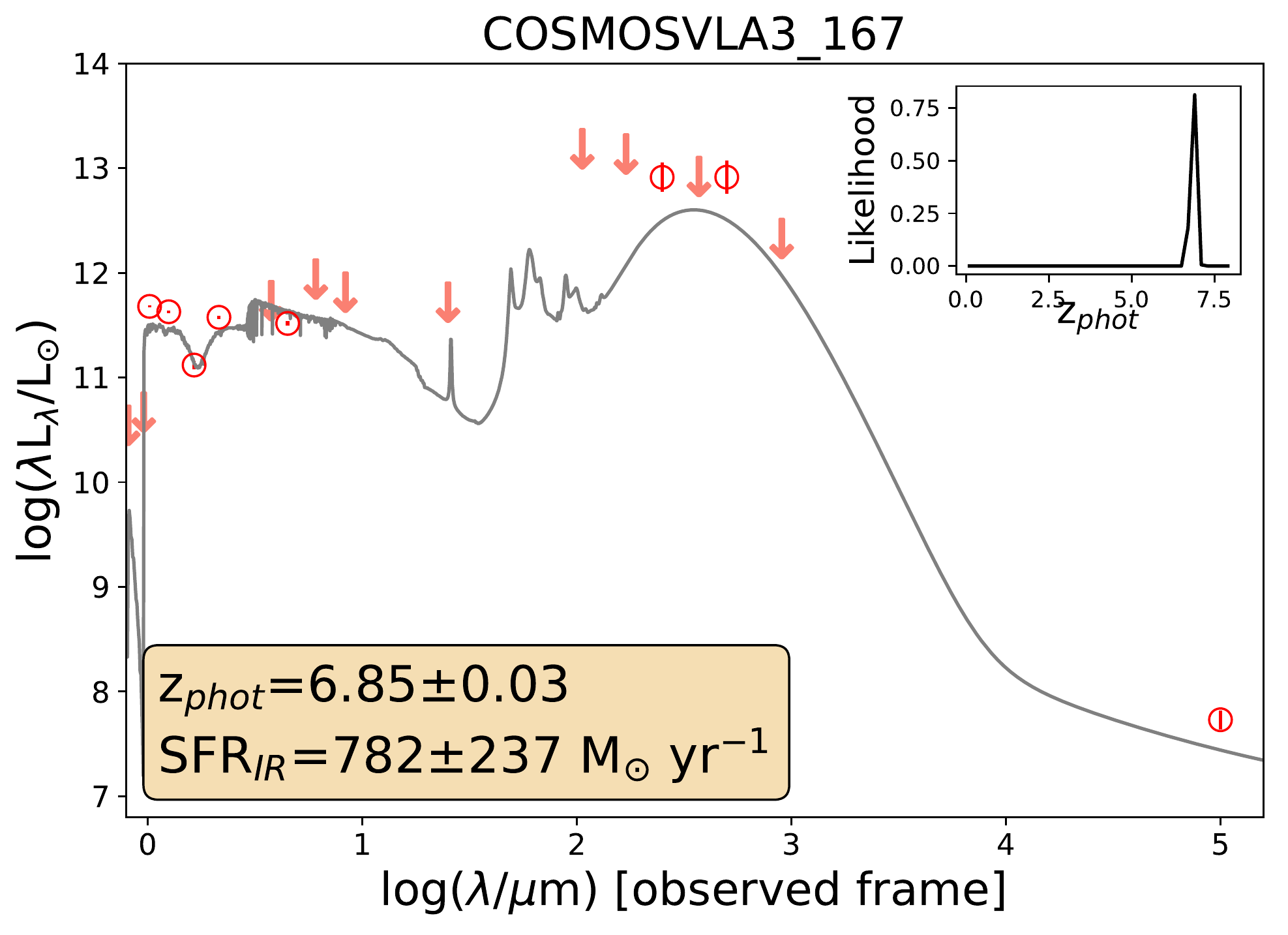}\includegraphics[scale=0.3]{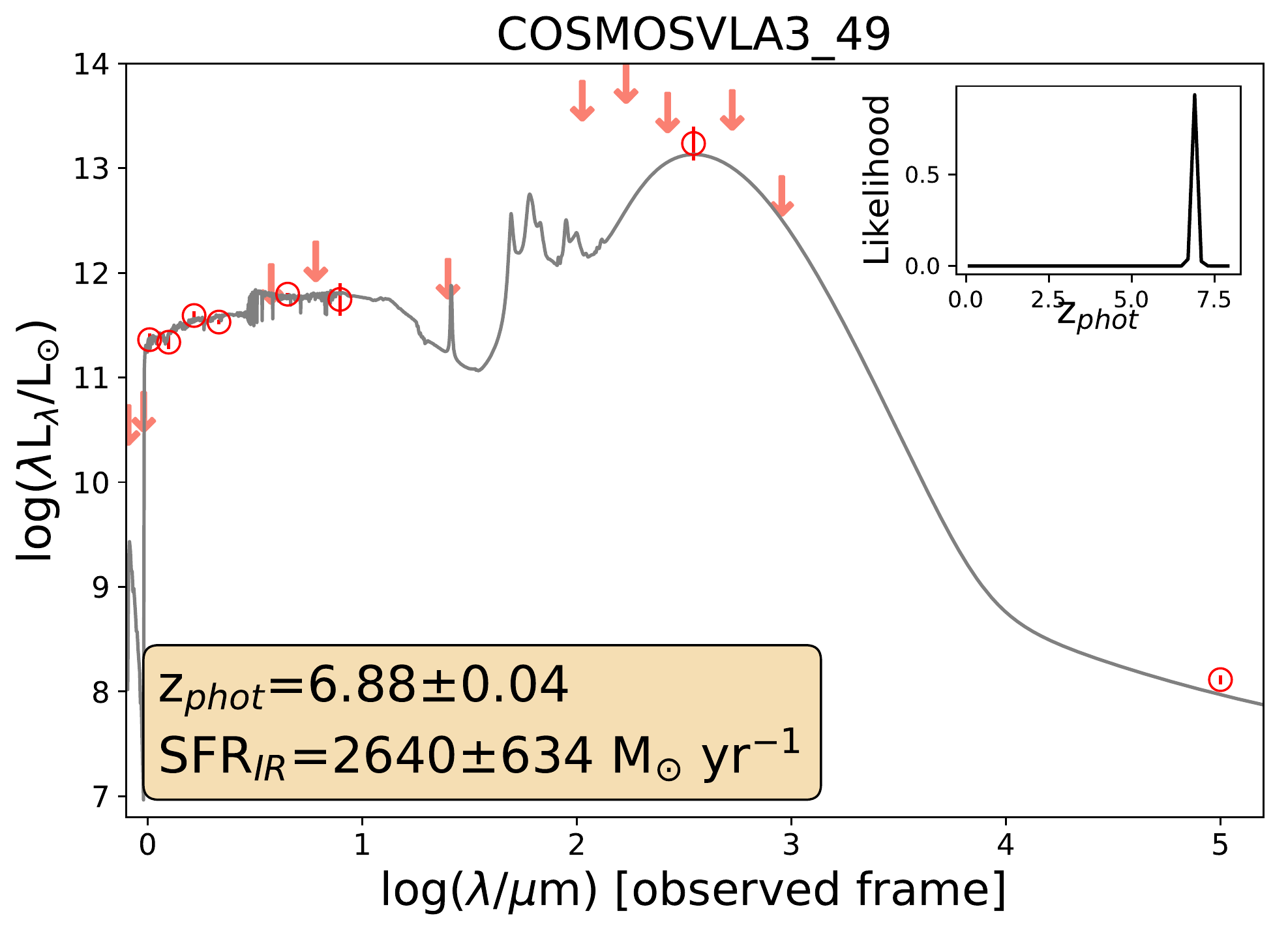}\includegraphics[scale=0.3]{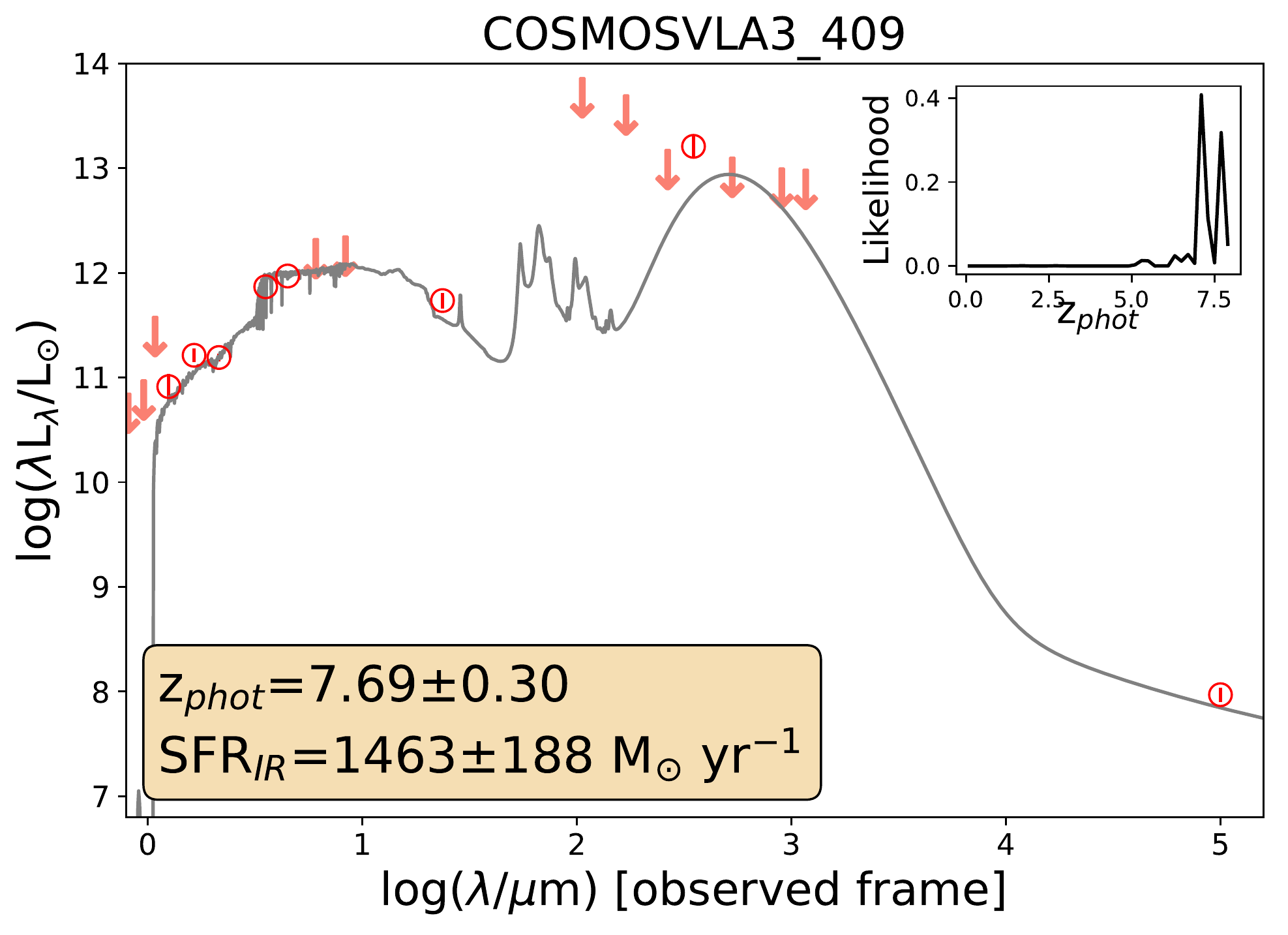}
	\includegraphics[scale=0.3]{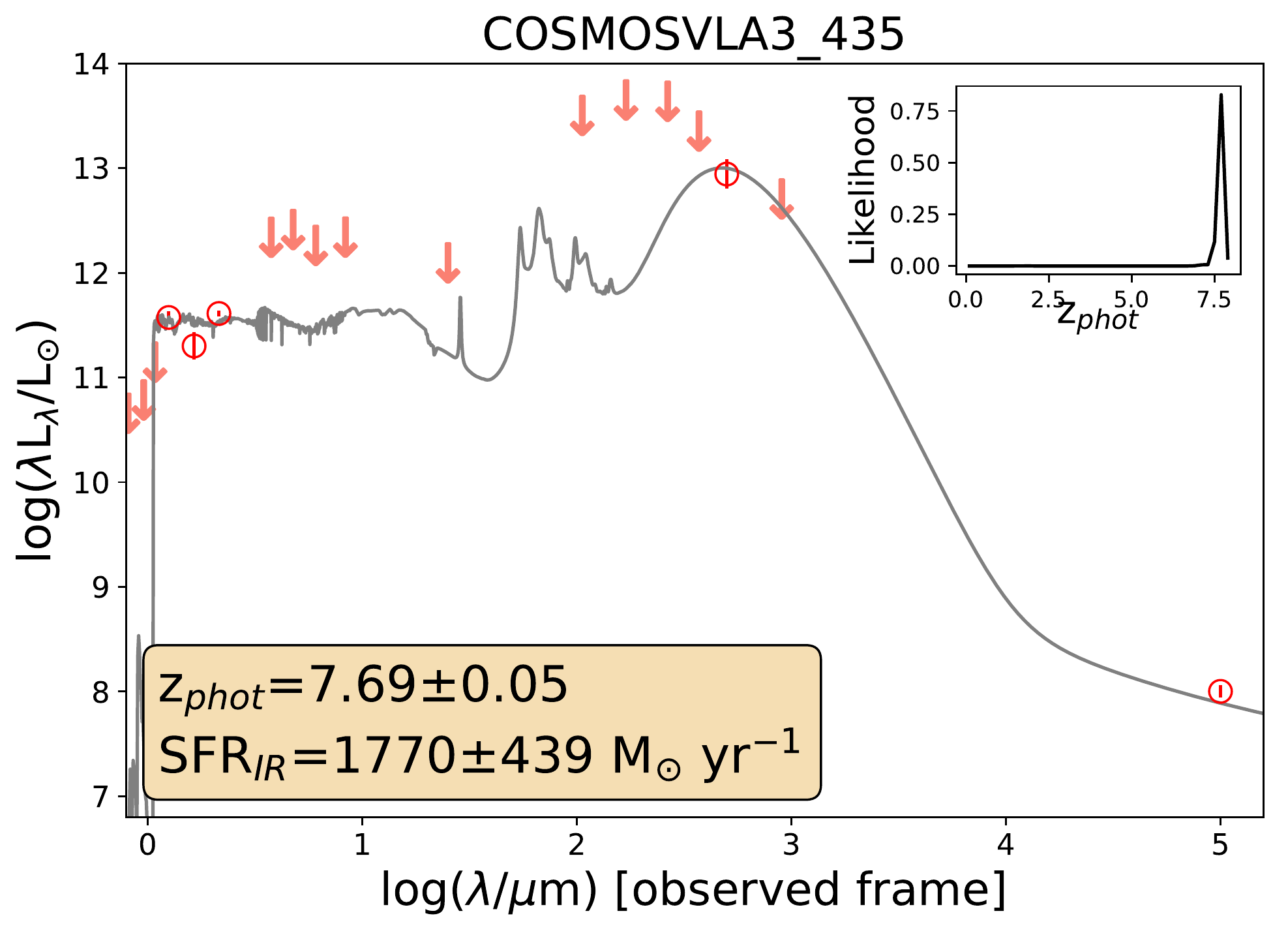}
	\caption{Continued from Fig. \ref{fig:highz}.}
	\label{fig:highz2}
\end{figure*}

\acknowledgments{
This paper is dedicated to the memory of Olivier Le F$\grave{\mathrm{e}}$vre.
We thank the anonymous referee for his/her useful suggestions to improve the paper. 
MT thanks Francesca Pozzi for useful discussions on dust temperature. 
MT, AC and MG acknowledge the support from grant PRIN MIUR 2017 20173ML3WW\_001.
We acknowledge the use of Python (v.2.7) libraries in the analysis.
This work is partly based on data products from observations made with ESO Telescopes at the La Silla Paranal Observatory under ESO programme ID 179.A-2005 and on data products produced by CALET and the Cambridge Astronomy Survey Unit on behalf of the UltraVISTA consortium.}

\appendix
\section{Comparison with the Wang et al. sample}

In the  \citet{wang2019} sample of H-dropouts in the COSMOS field, four out of 18 galaxies have a detected radio counterpart at 3 GHz. 
Two of them are below our chosen detection threshold (i.e. S/N$>$5.5). 
The remaining two are compatible with our own selection, although they are not included in the sample of 197 \emph{isolated} galaxies analyzed in this work. 
Therefore $\sim$11$\%$ of the Wang et al. sample is overlapping with our sample.
The Wang et al. selection is based on the H and IRAC2 (4.5 $\mu$m) bands. 
In particular, their galaxies are H-dropouts with a 5$\sigma$ limiting magnitude of $H < 27.4-27.8$ (in the COSMOS field), with a counterpart at 4.5 $\mu$m, IRAC2 $>$ 24.0.
With respect to the Wang et al. selection criteria, our sample of 197 radio-selected galaxies can be divided into three groups.

\begin{itemize}
\item \emph{Group 1}: galaxies which are detected in the H-band (43$\%$ of the sample). These galaxies are not consistent with the Wang et al. criteria, since they are not H-dropouts.
\item \emph{Group 2}: galaxies which are not detected neither in the H-band nor in the IRAC2 band (20$\%$ of the sample). These galaxies are excluded by the Wang et al. criteria, since they do not have an IRAC2 counterpart. We notice here that the IRAC limits are similar in the two works.
\item \emph{Group 3}: galaxies which are not detected in the H-band, but which have a counterpart in the IRAC2 band (37$\%$ of the sample). These galaxies could be potentially overlapping with Wang et al. criteria, because the 5$\sigma$ limiting magnitude in the H-band of the UltraVISTA DR4 maps is $H = 25.2-24.1$. 
\end{itemize}
In order to quantify the actual overlap between the radio-selected galaxies in \emph{Group 3} and the Wang et al. sample we performed two tests.
First, we stacked the galaxies in \emph{Group 3} in the H-band.
The measured median flux is $H = 26.57 \pm 0.23$, which is brighter than the limiting magnitude of the Wang et al. selection, meaning that at least 50$\%$ of \emph{Group 3} galaxies are not consistent with the Wang et al. selection.
As a second test we examined the best-fit magnitudes of \emph{Group 3} galaxies in the \emph{primary} sample: 70$\%$ of these galaxies have H $ < $ 27.4 mag, confirming the result from the first test.
By summing up all \emph{Group 1} and 2 and 50$\%$ of \emph{Group 3} galaxies we conclude that at least $\sim$82$\%$ of the UV-dark radio-selected galaxies are not consistent with the H-dropout selection by \citet{wang2019}.
We find a similar percentage (83$\%$) when focusing only on the common redshift range between the two samples (z$>$3).

\bibliographystyle{aasjournal}
\bibliography{references} 

\end{document}